\documentclass[singlecolumn,preprintnumbers,amsmath,amssymb,notitlepage,nofootinbib,longbibliography,superscriptaddress]{revtex4-1}

\usepackage{natbib}
\usepackage{graphicx}
\usepackage{dcolumn}
\usepackage{dsfont}
\usepackage{bm}
\usepackage{amsmath,amssymb,amsthm}
\usepackage{braket}
 \usepackage{cancel}
\usepackage[dvipsnames]{xcolor}
\usepackage{physics}

\usepackage{thmtools}
\usepackage{thm-restate}

\usepackage{letltxmacro}
\LetLtxMacro{\oldcite}{\cite}
\renewcommand{\cite}[1]{\mbox{\oldcite{#1}}}

\usepackage{algpseudocode,algorithm,algorithmicx}

\makeatletter
\renewcommand{\p@subsection}{}
\renewcommand{\p@subsubsection}{}
\makeatother

\usepackage{xcolor}
\usepackage{url}
\usepackage[colorlinks=true,breaklinks=true,allcolors=blue!60!black]{hyperref}

\usepackage{microtype}
\usepackage{soul}

\usepackage{enumerate}

\newcommand{\gen}{\textsf{GEN}}
\renewcommand{\eval}{\textsf{EVAL}}

\newcommand{\lqc}{\mathrm{LQC}}
\newcommand{\sample}{\mathrm{SAMPLE}}
\newcommand{\query}{\mathrm{query}}
\newcommand{\SQ}{\mathrm{SQ}}
\newcommand{\pr}{\mathrm{Pr}}
\newcommand{\tv}{\mathrm{d}_\mathrm{TV}}
\newcommand{\A}{\mathcal{A}}
\newcommand{\btheta}{\boldsymbol{\theta}}
\newcommand{\bphi}{\boldsymbol{\phi}}

\newtheorem{result}{Result}

\newtheorem{corollary}{Corollary}

\newtheorem{observation}{Observation}
\newtheorem{conjecture}{Conjecture}
\newtheorem{definition}{Definition}

\newtheorem{remark}{Remark}

\declaretheorem{theorem}
\declaretheorem{lemma}

\declaretheorem[numbered=no,
  name=
  Problems: PAC probabilistic modelling of quantum circuit Born machines]{PACBorn}

\makeatletter 
\hypersetup{pdftitle = {Learnability of the output distributions of local quantum circuits},
    pdfauthor = {
        Marcel Hinsche, 
        Marios Ioannou, 
        Alex Nietner, 
        Jonas Haferkamp, 
        Yihui Quek, 
        Dominik Hangleiter, 
        Jean-Pierre Seifert, 
        Jens Eisert, 
        Ryan Sweke},
    pdfsubject = {
        Machine learning, 
        quantum computation},
    pdfkeywords = {
        quantum machine learning, 
        PAC learning, 
        quantum advantage, 
        Born machine,
        quantum circuit, 
        Clifford circuits,
        statistical query model, 
        generative modelling, 
        randomized statistical dimension,
        property testing
         }
      }
\makeatother

\begin{document}

\title{Learnability of the output distributions of local quantum circuits}

\newcommand{\fu}{Dahlem Center for Complex Quantum Systems, Freie Universit{\"a}t Berlin, 14195 Berlin, Germany}
\newcommand{\stan}{Information Systems Laboratory, Stanford University, Stanford, CA  94305, USA}
\newcommand{\tu}{Electrical Engineering and Computer Science,
TU Berlin, D-10587 Berlin, Germany}
\newcommand{\mary}{Joint Center for Quantum Information and Computer Science (QuICS), University of Maryland and NIST, College Park, MD 20742, USA}
\newcommand{\hzb}{Helmholtz-Zentrum Berlin f{\"u}r Materialien und Energie, 14109 Berlin, Germany}
\newcommand{\hhi}{Fraunhofer Heinrich Hertz Institute, 10587 Berlin, Germany}
\newcommand{\sit}{FhG SIT, D-64295 Darmstadt, Germany}

\author{M.~Hinsche}
\address{\fu} 
\author{M.~Ioannou}
\address{\fu} 
\author{A.~Nietner}
\address{\fu} 
\author{J.~Haferkamp}
\address{\fu} 
\address{\hzb}
\author{Y.~Quek}
\address{\stan}
\address{\fu}
\author{D.~Hangleiter} 
\address{\mary}
\address{\fu}
\author{J.-P.~Seifert}
\address{\tu}
\address{\sit}
\author{J.~Eisert}
\address{\fu}
\address{\hzb}
\address{\hhi}
\author{R.~Sweke}
\address{\fu}

\begin{abstract}
There is currently a large interest in understanding the potential advantages quantum devices can offer for probabilistic modelling. In this work we investigate, within two different oracle models, the probably approximately correct (PAC) learnability of quantum circuit Born machines, i.e., the output distributions of local quantum circuits. We first show a negative result, namely, that the output distributions of super-logarithmic depth Clifford circuits are not sample-efficiently learnable in the statistical query model, i.e., when given query access to empirical expectation values of bounded functions over the sample space. This immediately implies the hardness, for both quantum and classical algorithms, of learning from statistical queries the output distributions of local quantum circuits using any gate set which includes the Clifford group. 
As many practical generative modelling algorithms use statistical queries -- including those for training quantum circuit Born machines -- our result is broadly applicable and strongly limits the possibility of a meaningful quantum advantage for learning the output distributions of local quantum circuits.
As a positive result, we show that in a more powerful oracle model, namely when directly given access to samples, the output distributions of local Clifford circuits are computationally efficiently PAC learnable by a classical learner. 
Our results are equally applicable to the problems of learning an algorithm for generating samples from the target distribution (generative modelling) and learning an algorithm for evaluating its probabilities (density modelling). 
They provide the first rigorous insights into the learnability of output distributions of local quantum circuits from the probabilistic modelling perspective.
\end{abstract}

\maketitle

\section{Introduction}\label{s:intro}

A large body of recent work has tried to identify concrete machine learning (ML) tasks for which quantum machine learning (QML) methods could demonstrate a well-defined and meaningful advantage over classical methods. In particular, it is known that if one allows finely tuned and highly structured datasets, as well as special purpose quantum learning algorithms (i.e., learners designed specifically for the finely tuned task), then there do indeed exist problems for which quantum learners can obtain meaningful advantages \cite{arunachalam2017guest,liu2021rigorous, Sweke2021quantumversus}. However, ideally one would like to demonstrate that one can obtain an advantage for 
practically relevant problems, using ``generic" quantum learning algorithms, preferably those which can be executed on near-term devices in the hybrid quantum classical framework~\cite{bharti2021noisy,cerezo2020variational,benedetti2019parameterized}. While a large proportion of recent QML research has been focused on supervised learning, one area that has seemed particularly promising for demonstrating such quantum/classical separations is unsupervised generative modelling. 

In an unsupervised generative modelling problem, one is given some type of oracle access to the unknown target distribution. The goal of the learning algorithm is to output, with high probability, an approximate \textit{generator} for the target distribution -- i.e., an algorithm for generating samples from some distribution which is sufficiently close to the target distribution~\cite{Sweke2021quantumversus, Kearns:1994:LDD:195058.195155}. Many highly relevant practical ML problems are of this type, and as such the development and application of classical methods for this problem -- such as \emph{generative adversarial networks (GANs)}~\cite{goodfellow2014generative}, \emph{variational auto-encoders}~\cite{kingma2013auto} and \emph{normalizing flows}~\cite{Kobyzev_2020} -- is a highly active research topic. Given this, the development of quantum models and algorithms for generative modelling is of natural interest and a variety of approaches, such as 
\emph{quantum circuit Born machines (QCBMs)}~\cite{coyle2020born, Liu_2018, Benedetti_2019, Gaoeaat9004}, \emph{quantum GANs}~\cite{Dallaire_Demers_2018,Hueaav2761, lloyd2018quantum, chakrabarti2019quantum} and \emph{quantum Hamiltonian-based models}~\cite{verdon2019quantum} have also been proposed and implemented \cite{rudolph2020generation}. 

QCBMs are a particularly promising class of models, which are based on the simple observation that measuring the output state vector~$|\psi\rangle = U|0\ldots 0\rangle$ of a quantum circuit $U$, in the computational basis, provides a sample from the ``Born distribution" $P_{U}$ defined by the circuit, i.e., the distribution over bit strings for which 
\begin{equation}
    P_{U}(x) := |\langle x| \psi\rangle|^2 = |\langle x|U|0\ldots 0\rangle|^2.
\end{equation}
Given this observation, QCBM based generative modelling algorithms typically work by iteratively updating the parameters $\boldsymbol{\theta}$ of a parameterized quantum circuit $U(\boldsymbol{\theta})$, until the Born distribution of the circuit matches as closely as possible -- with respect to some loss function -- the unknown target distribution \cite{coyle2020born,Liu_2018,Benedetti_2019,Gaoeaat9004}.

In light of the known hardness of classically simulating certain classes of local quantum circuits~\cite{Bremner_2010,terhal2002adaptive,aaronson2016complexitytheoretic}, some recent works have conjectured or provided numerical evidence for the classical hardness of the generative modelling problem associated with QCBMs ~\cite{coyle2020born,niu2020learnability}. 
More specifically, these works have suggested that learning a generator for the Born distributions of local quantum circuits, when given access to samples from such distributions, may be hard for classical learning algorithms. In contrast, it seems natural to conjecture that this particular generative modelling problem is computationally feasible for QCBM based learning algorithms. 
This is because such algorithms naturally use parameterized local quantum circuits as generators by construction, and need only to identify the correct circuit parameters. In particular, a separation between the power of quantum and classical generative modelling algorithms has already been established, using a highly fine tuned concept class and learning algorithm~\cite{Sweke2021quantumversus}. However, the hope has been that one could demonstrate a similar separation using a generic QCBM based learner by considering the learnabilility of QCBM distributions themselves. Moreover, as quantum circuit Born machines are known to be highly expressive \cite{glasser2019expressive}, the hope has been that such a quantum/classical separation might translate to a quantum advantage for practical generative modelling problems. More specifically, when making the decision to use a QCBM for a practical probabilistic modelling problem, one is making the implicit assumption that the target distribution can indeed be well approximated by the Born distribution of some local quantum circuit. It is therefore well motivated to try prove a separation between the power of QCBM based algorithms and classical algorithms for learning the output distributions of QCBMs themselves.
However, in order to demonstrate such a separation via QCBMs one requires two results: Firstly, a rigorous proof of the classical hardness of the generative modelling problem associated with the output distributions of local quantum circuits, and secondly, a rigorous proof of the efficiency of QCBM based algorithms for the same task. 

\subsection{Overview of this work}\label{ss:overview}

Motivated by these questions, we study in this work the learnability of the output distributions of local quantum circuits within the probably approximately correct (PAC) framework for probabilistic modelling~\cite{Kearns:1994:LDD:195058.195155,Sweke2021quantumversus}. Since its introduction, Valiant's model of PAC learning \cite{valiant1984theory}, along with a variety of natural extensions and modifications, has provided a fruitful framework for studying both the computational and statistical aspects of machine learning \cite{kearns1994introduction,shalev2014understanding}, and for the rigorous comparison of quantum and classical learning algorithms \cite{arunachalam2017guest}. In addition to providing results for \textit{generative modelling}, we also study the related problem of \textit{density modelling}. 
In this setting, the goal of the learner is not to generate new samples from the target distribution, but to output, with high probability, a sufficiently accurate algorithm for evaluating the probabilities of events with respect to the target distribution -- i.e. an algorithm which when given an event $x$ outputs the associated probability $P(x)$. We refer to such an algorithm as an \textit{evaluator} for the target distribution. 

Moreover, we study both of these probabilistic modelling problems with respect to two different models of access to the unknown target distribution. The first model we call the \textit{sample model}, as we assume in this model that the learner has access to a \textit{sample oracle} which provides samples from the unknown target distribution. The second model is the \textit{statistical query (SQ) model}, which has originally been introduced by Kearns in Ref.~\cite{kearns1998efficient} as a natural restriction of the sample model, and which in the context of supervised learning of Boolean functions,  guarantees noise-robustness of the associated learning algorithm. In the SQ model, learners do not have access to samples from the target distribution, but only to approximate averaged statistical properties of the unknown target distribution. More specifically, learners have access to an \textit{SQ oracle}, which when queried with a function, provides an approximation to the expectation value of the output of that function, with respect to inputs drawn from the unknown target distribution. Since the SQ model is a strict restriction of the sample model, hardness of learning in the SQ model \textit{does not} imply hardness of learning in the sample model. Still, within the probabilistic modelling context, hardness results in the SQ model are of interest for two important reasons. Firstly, the SQ model provides a natural way to restrict one's attention to learning algorithms which, if given access to a sample oracle, always use their samples from the target distribution to calculate approximate expectation values of functions via sample mean estimates. As we will show, many \textit{generic} implicit generative modelling algorithms -- i.e. those which are not designed to exploit a particular structure in the target distribution class -- are of this type, including those for training quantum circuit Born machines \cite{Liu_2018,coyle2020born,mohamed2017learning}. As such, hardness results in the SQ model apply to many implicit generative modelling algorithms of practical interest, and in particular to those which are often used for the concept class of interest in this work. Secondly, while it is often easier to obtain lower bounds on the query complexity of learning algorithms in the SQ model -- via constructive quantities known as \textit{statistical dimensions} -- there are very few examples of learning problems which are known to be hard in the SQ model, but easy in the sample model \cite{Feldman2016,feldman2017general}. As such, hardness in the SQ model is often taken as strong evidence for hardness in the sample model. 

In summary, we study in this work the following problems, which are stated more formally in Section \ref{s:prelim}:

\begin{PACBorn}[informal]
Let $\mathcal{C}$ be the set of output distributions corresponding to a class of local quantum circuits. Given either sample-oracle or SQ-oracle access to some unknown distribution $P\in\mathcal{C}$,  output, with high probability, either
\begin{description}
    \item[$\qquad$ generative modelling] an efficient generator, or
    \item[$\qquad$ density modelling] an efficient evaluator
\end{description}
for a distribution $\tilde{P}$ which is sufficiently close to $P$.
\end{PACBorn}
If there exists either a sample or computationally efficient algorithm which, with respect to either the sample oracle or the SQ oracle, solves the generative (density) modelling problem associated with a given set of distributions $\mathcal{C}$, then we say that $\mathcal{C}$ is sample or computationally efficiently generator (evaluator) learnable within the relevant oracle model. We are particularly interested in this work in establishing the existence or non-existence, of efficient quantum or classical learning algorithms, for the output distributions of various classes of local quantum circuits, within both the sample and statistical query model.

\subsection{Main results}
Given this motivation and context, we provide two main results, which stated informally, are as follows:

\begin{result}[Informal version of Corollary \ref{c:no_sample_efficient_generic}]\label{r:informal_hardness}
The concept class consisting of the output distributions of super-logarithmic depth nearest neighbour Clifford circuits is not sample efficiently PAC generator-learnable or evaluator-learnable, in the statistical query model.
\end{result}

\begin{result}[Informal version of Theorem \ref{t:clifford_learn}]\label{r:informal_learnability}
The concept class consisting of the output distributions of nearest neighbour Clifford circuits is both sample and computationally efficiently classically PAC generator-learnable and evaluator-learnable, in the sample model. 
\end{result}

These results provide some first concrete insights into the learnability of the output distributions of local quantum circuits from a probabilistic modelling perspective, and are of interest for a variety of reasons. Firstly, we note that Result \ref{r:informal_hardness} applies not just to Clifford circuits: it implies the hardness of learning the output distributions of any nearest neighbour quantum circuit whose gates come from some gate set which includes the two-qubit Clifford group. However, we choose to stress the special case of local Clifford circuits in our statement of Result \ref{r:informal_hardness}, as it allows us to highlight the fact that the generative modelling problem associated with a class of local quantum circuits can be hard, even when the class of circuits are efficiently classically simulatable! More specifically, local Clifford circuits are known to be classically efficiently simulatable, in the sense that given a description of the quantum circuit, there exist classically efficient algorithms both to evaluate the probabilities of events, and to sample from the associated Born distribution~\cite{gottesman1998heisenberg,Aaronson_2004}. As such, while the probabilistic modelling problems we consider are naturally analogous to classical simulation problems -- but with SQ access to the distribution as input rather than a circuit description -- our first result establishes that learning both generators and evaluators for the output distribution of a local quantum circuit from SQ queries can be hard, even when outputting a generator or an evaluator from a circuit description can be done efficiently.

Secondly, we stress that as Result \ref{r:informal_hardness} provides a \textit{query complexity} lower bound, it holds \textit{for both quantum and classical learners}. As such, this result directly implies that, at least in the statistical query model, one \textit{cannot} use the concept class of local quantum circuit output distributions to demonstrate a meaningful separation between the power of quantum and classical generative modelling algorithms. More specifically, as mentioned before, any such separation requires both a classical hardness result -- i.e. a proof that a given concept class is not efficiently learnable via classical learning algorithms -- and a quantum learnability result -- i.e. an explicit efficient quantum learning algorithm for the given concept class.  However, our work establishes that, at least in the SQ model, efficient quantum learnability of the output distributions of (super-logarithmically deep) local quantum circuits is not possible, \textit{even for classically simulatable circuit classes}. This result therefore provides a direct obstacle to the goal of proving an exponential quantum advantage for generative modelling via QCBMs as (a) learning algorithms for QCBMs typically use statistical queries and (b) the concept class of output distributions of local quantum circuits is certainly the most natural set of distributions with which to try prove an advantage for QCBMs. 

Additionally, as mentioned before, hardness results in the SQ model are often taken as strong evidence for computational hardness in the sample model. However, as Result \ref{r:informal_learnability} covers all local Clifford circuits, and in particular those of super-logarithmic depth, we see that the distribution concept class of Result \ref{r:informal_hardness} provides an interesting example of a generative modelling problem which is hard in the SQ model, but computationally efficient in the sample model. It is important however to stress that, in order to exploit individual samples from the target distribution, the efficient learning algorithm implied by Result \ref{r:informal_learnability} relies heavily on knowledge of the algebraic structure of stabilizer states (the output states of Clifford circuits). As such it remains an interesting open problem to understand whether the output distributions of more generic local quantum circuits are also learnable in the sample model, despite being hard to learn in the SQ model. 

Finally, we stress that while our work provides some first concrete insights into the learnability of the output distributions of local quantum circuits, there remain a variety of interesting open questions. In particular, there are many combinations of circuit depth, gate-type, oracle model, and learner-type which are not covered by our results. In light of this, we provide in Section \ref{s:discussion} a detailed description of some of the open questions prompted by this work, along with multiple explicit conjectures.

\subsection{Proof Techniques}

Finally, before proceeding we mention briefly some of the proof techniques involved in establishing our results. For Result \ref{r:informal_hardness}, we exploit the fundamental conceptual observation from property testing, which is that testing properties of an object can sometimes be easier than learning an object, and as such one can often lower bound the query complexity of a learning problem by lower bounding the query complexity of a suitable property testing problem~\cite{canonne2020survey,goldreich2017introduction}. In our case, we observe that lower bounds on the query complexity of identity testing, with the additional promise that the unknown distribution is from the concept class to be learned, is sufficient to prove query complexity lower bounds for the probabilistic modelling problems we are interested in. By phrasing the problem of identity testing with a promise as a \textit{decision problem} (as defined in Ref. \cite{feldman2017general}), we are then able to exploit existing results from Feldman \cite{feldman2017general}, who has shown that the query complexity of any decision problem in the SQ model can be completely characterized by the \textit{randomized statistical dimension} of the problem. As such, our main technical result is a lower bound for the randomized statistical dimension of a suitably constructed decision problem. Specifically, this decision problem encodes the problem of testing the identity of some distribution, which is promised to be the output distribution of a local quantum circuit. In order to obtain this lower bound, we rely on techniques for calculating moments over the unitary group \cite{low2010pseudorandomness,hunterjones2019unitary,barak2021spoofing,dalzell2020random}. For Result \ref{r:informal_learnability}, we exploit the known relationship between stabilizer states -- the output states of Clifford circuits -- and affine subspaces of $\mathbb{F}^n_2$ (the $n$-dimensional vector space over the finite field of two elements). More specifically, we use the observation that the Born distribution of any Clifford circuit is the uniform distribution over some affine subspace of $\mathbb{F}^n_2$~\cite{Dehaene_2003,montanaro2017learning}. We then show that one can efficiently recover a description of an affine subspace of $\mathbb{F}^n_2$ when given samples from the uniform distribution over that space.

\subsection{Structure of this work}

\noindent This work is structured as follows: We begin by discussing in Section \ref{s:related_work} the relation of our work to existing work. In particular, we use this section to stress the distinctions between the problem we consider and a variety of related problems in quantum computing and computational learning theory. In particular, we wish to make clear how the results we obtain here do not follow immediately from known results in related areas. With this context in hand we then introduce formally in Section \ref{s:prelim} all the preliminaries necessary for this work. In particular, we define the PAC model for distribution learning, the distribution concept class of local quantum circuits, decision problems in the SQ model, and fundamental linear algebra over $\mathbb{F}^n_2$. These preliminaries allow us to present Result \ref{r:informal_hardness} in Section \ref{s:main_results} and Result \ref{r:informal_learnability} in Section \ref{s:clifford_learnability}. Finally we conclude in Section \ref{s:discussion} with a discussion and an explicit list of open questions and conjectures.

\section{Relation to existing work}\label{s:related_work}

\begin{figure}
\centering
\includegraphics[width=1.0\linewidth]{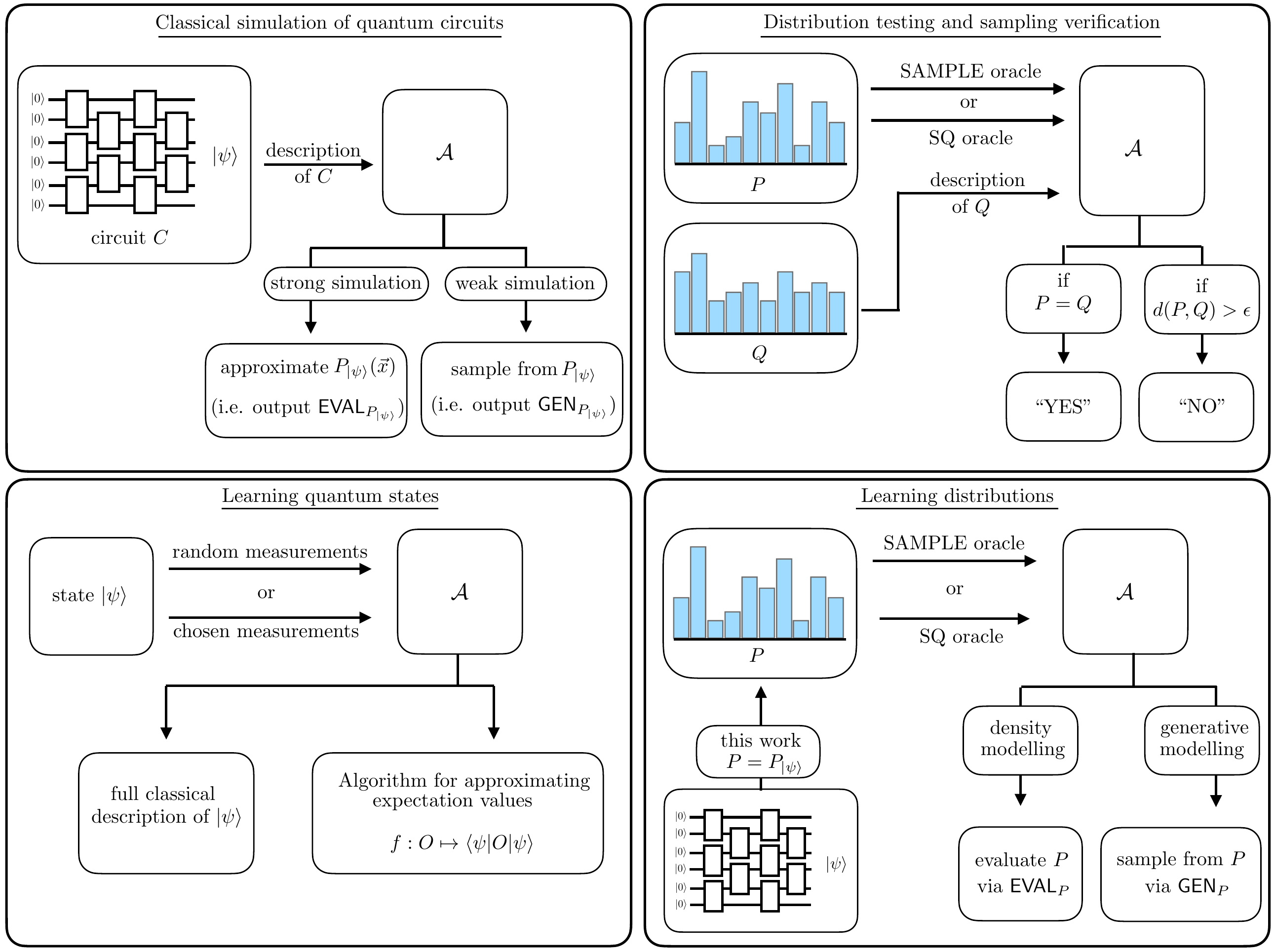}
\caption{An illustration of the related computational and learning problems discussed in Section \ref{s:related_work}.}\label{f:related_work}
\end{figure}

The probabilistic modelling problems we study in this work -- defined informally in Section \ref{ss:overview} -- are closely related to, but distinct from, a variety of different computational problems in quantum information and computational learning theory. In order to make these distinctions clear, and to clarify the extent of some potential reductions between learnability results for probabilistic modelling and known results in related areas, we provide here a brief and informal discussion of these related problems. These related problems are also illustrated in Figure \ref{f:related_work}.\newline

\noindent\textbf{Classical simulation of quantum circuits:} Given a specific class of quantum circuits, it is of fundamental interest to understand whether, and in which sense, quantum circuits from the given class are efficiently classically simulatable. Typically one differentiates between the notions of \textit{weak} classical simulation and \textit{strong} classical simulation. As illustrated in Figure~\ref{f:related_work}, in both instances one is given as input an efficient classical description of the quantum circuit. Using the language of probabilistic modelling, in a weak classical simulation the task is to output a generator for the Born distribution of the output state of the quantum circuit, while for strong classical simulation the task is to output an evaluator for the Born distribution of the circuit. For both strong and weak simulation, if there exists an efficient algorithm which can succeed for all circuits in the specific class, then one says the class of circuits is either weakly or strongly efficiently classical simulatable. If one can prove that no such efficient classical algorithm exists, then one says that classically simulating the given circuit class is worst-case hard.  Alternatively, if one can prove that, with high probability when drawing a circuit randomly from the class, the simulation task cannot be performed efficiently, then one says that simulating the given class of circuits is average-case hard. We stress that while the desired \textit{output} of a strong/weak classical simulation is the same as the desired output of the associated density/generative modelling problem, the inputs differ significantly. More specifically, in the case of classical simulation one is given a description of the circuit as input, while in the probabilistic modelling setting we are concerned with here, one is given only some sort of oracle access to the Born distribution of the output state of the circuit. Moreover, while there is currently a large interest in establishing the average-case hardness of weak classical simulation for certain classes of local quantum circuits~\cite{bouland2019complexity,movassagh_cayley_2019,bouland_noise_2021}, typically in the probabilistic modelling setting one is concerned with establishing worst-case hardness. We note that while multiple previous works have conjectured implications between the hardness of weak-classical simulation of quantum circuits and the hardness of the associated generative modelling problem~\cite{Liu_2018,coyle2020born}, our results establish firmly that, at least in the statistical query model, the generative modelling problem associated with a class of local quantum circuits can be computationally hard, even when the weak-classical simulation problem is easy. As such, despite what has been previously suggested,, one \textit{cannot} straightforwardly use the hardness of a classical simulation for a given class of quantum circuits to prove the hardness of the associated generative modelling problem. Indeed, as stressed above, these are completely different computational problems. \newline

\noindent\textbf{Distribution testing and verification of quantum circuit sampling:} The field of distribution testing is concerned with the development of algorithms for testing whether or not an unknown probability distribution has a given property \cite{canonne2020survey}. Given that \textit{learning} a complete description of a distribution often allows one to test properties of the distribution, lower bounds on the query complexity of testing algorithms often imply lower bounds on the query complexity of learning algorithms \cite{goldreich2017introduction}. One particularly important distribution testing problem is that of \textit{identity testing}: Given a complete description of some known distribution $Q$, as well as some type of oracle access to an unknown distribution $P$, decide whether $P=Q$ or is at least a certain distance away. Using optimality results for distribution identity testing \cite{
valiant2017automatic}, one can show that, for certain classes of local quantum circuits, there exists no sample-efficient classical algorithm for testing, from samples, whether or not the samples are from the Born distribution of a given quantum circuit~\cite{hangleiter2019sample,hangleiter2020sampling}. Using the standard intuition from property testing -- namely that learning algorithms often imply testing algorithms --
one might think that the existence of a sample-efficient algorithm for \textit{learning} a generator for the Born distributions of local quantum circuits would imply the existence of a sample-efficient algorithm for \textit{testing} whether or not samples, which come from some generator, are indeed coming from the Born distribution of a given local quantum circuit. Indeed, if this was the case, then one could use the known hardness results for testing the Born distributions of local quantum circuits to rule out the existence of efficient generator-learning algorithms for the same class of circuits. Unfortunately
-- and interestingly -- however, this is not the case. One can show that in order to obtain hardness of learning results, one requires hardness results not for the standard distribution identity testing problem, but rather for the problem of distribution identity testing with the additional promise that the samples to be tested are coming from some distribution in the concept class of the learner. In this more restricted problem, the distribution has to be distinguished from fewer distributions as compared to the unrestricted identity testing problem. It is this fact which motivates our reduction between generator-learning and a specific \textit{decision problem} \cite{feldman2017general}, which as explained in Section~\ref{ss:SQ} can be viewed precisely as distribution identity testing with an additional promise. \newline

\noindent\textbf{Learning quantum states:} There exist a wide variety of different notions of what it means to ``learn a quantum state". Perhaps most intuitive is that of quantum state tomography, in which given the ability to perform arbitrary efficient measurements on multiple identical copies of an unknown state, one would like to learn a full classical description of the state \cite{BenchmarkingReview}. As obtaining a full classical description of a quantum state, in the general case, precludes efficient algorithms, multiple refinements of quantum state tomography have been introduced, in which the goal is only to predict some properties of the unknown quantum state, such as expectation values of particular observables. Examples of such refinements include Aaronson's extension of the PAC framework for quantum states \cite{Aaronson_2007}, Aaronson's shadow tomography framework \cite{aaronson2019shadow} and classical shadow learning \cite{Huang_2020}. In the case when the quantum state to be learned is the output state of a local quantum circuit -- as for example in previous works on learning stabilizer states~\cite{Rochhetto,gollakota2021hardness} -- the above mentioned state-learning problems are similar in some respects to the probabilistic modelling problem we consider here, while differing in a few essential ways. Most importantly, in the state learning setting one typically has access to the outcomes of a variety of different types of measurements, where as in the probabilistic modelling setting one only has oracle access to the Born distribution of the unknown state - i.e. to the outcomes of measurements in the computational basis. Similarly, in the probabilistic modelling setting we are only concerned with obtaining either a generator or evaluator for the Born distribution of the state, as opposed to either a full classical description of the quantum state, or an algorithm for predicting the expectation values of different observables.\newline

\noindent\textbf{Distribution learning:}  Given the fundamental importance of probabilistic modelling for a wide variety of applications, there is by now a large body of results on the learnability of different classes of probability distributions~\cite{Kearns:1994:LDD:195058.195155, canonne2020short,diakonikolas2016learning,kamath2015learning,de2014learning}. While the majority of such work has been in the sample oracle model, recent work has also started to explore such questions in the statistical query model \cite{diakonikolas2017statistical}. Up until now however, there has been no work on the learnability of Born distributions of local quantum circuits. As such, while we rely on similar techniques to previous works on probabilistic modelling in the PAC framework -- namely reductions from property testing and lower bounds via statistical dimensions -- our work is distinct by virtue of the class of distributions we consider, which is motivated by the desire to understand the potential advantages quantum probabilistic modelling algorithms may offer over classical approaches. \newline

\section{Preliminaries}\label{s:prelim}

\noindent We denote the set of all distributions over $\{0,1\}^n$ as $\mathcal{D}_n$. We denote the uniform distribution over $\{0,1\}^n$ as $\mathcal{U}_n$. If $n$ is implicit from the context, or not important for an argument, we will often omit the subscript. We will often consider subsets, $\mathcal{C}\subseteq\mathcal{D}_n$ which we refer to as \textit{distribution concept classes}. Given some distribution concept class $\mathcal{C}\subseteq\mathcal{D}_n$, a reference distribution $P\in \mathcal{C}$ and some $\epsilon > 0$, we use $\mathcal{B}(P,\epsilon)$ to denote the epsilon ball, with respect to the total variation distance,
around $P$ in $\mathcal{C}$, i.e.
\begin{equation}
    \mathcal{B}_{\mathcal{C}}(P,\epsilon) = \{D\in\mathcal{C}\,|\,\tv(D,P) < \epsilon\}.
\end{equation}
Once again, when $\mathcal{C}$ is clear from the context the subscript will be omitted. We denote the set of all probability measures over a set $X$ by $S^X$. We will use the notation $\mathcal{U}_X$ to denote the uniform measure over a set $X$, but for convenience we will often use the shorthand $x\sim X$ to denote $x$ sampled from $\mathcal{U}_X$. Given some oracle $O$, and a randomized algorithm $\A$, we will use the notation $\A^{O}$ to mean $\A$ with query access to $O$. We denote the unitary group of degree $n$ by $\mathrm{U}(n)$. Finally, as expectation values of function outputs with respect to randomly drawn inputs are a central aspect of this work, we define the following shorthand notation, which is used frequently:

\begin{definition}[Expectation values of function outputs]Given some function $\phi:\{0,1\}^n\rightarrow[-1,1]$, as well as some $P\in\mathcal{D}_n$, we use the notation $P[\phi]$ to denote the expectation value $\mathbb{E}_{x\sim P}[\phi(x)]$, i.e. 
\begin{equation}
    P[\phi] := \mathbb{E}_{x\sim P}[\phi(x)].
\end{equation}
\end{definition}

\subsection{PAC framework for probabilistic modelling}\label{ss:PAC_dist}

We formalize in this section the PAC framework for probabilistic modelling, building on and refining the definitions from Refs.~\cite{Kearns:1994:LDD:195058.195155,coyle2020born,Sweke2021quantumversus}. In order to build such a framework, the first thing we require is a meaningful notion of ``access to a distribution". We achieve this via the following oracles:

\begin{definition}[Distribution oracles]\label{d:oracles}
Given $P\in\mathcal{D}_n$ we define the sample oracle $\sample(P)$ as the oracle which, when queried, provides a sample from $P$. We denote this via
\begin{equation}
    \query[\sample(P)] = x\sim P.
\end{equation}
Additionally, given some $\tau \in [0,1]$, we define the statistical query oracle $\SQ_{\tau}(P)$ as the oracle which, when queried via some efficiently computable function $\phi:\{0,1\}^n\rightarrow [-1,1]$, responds with some $v$ such that $|P[\phi] - v| \leq \tau$. We denote this via
\begin{equation}
    \query[\SQ_{\tau}(P)](\phi) = v \text{ such that } |P[\phi] - v| \leq \tau.
\end{equation}
\end{definition}
We stress that for any distribution $P$, the oracle $\SQ_{\tau}(P)$ is specified via a tolerance parameter $\tau>0$, which determines the accuracy of the expectation values provided by $\SQ_{\tau}(P)$. In particular, we note that for any $\tau$ which decays at most inverse polynomially in $n$ -- i.e. $\tau = \Omega(1/\mathrm{poly}(n))$ -- one can 
straightforwardly use access to $\sample(P)$ to efficiently simulate access to $\mathrm{SQ}_{\tau}(P)$. Specifically, given any appropriate efficiently computable function $\phi$, one simply outputs the sample mean of the output of $\phi$ on polynomially many samples drawn from $\sample(P)$ \cite{Feldman2016,kearns1998efficient}. In light of this, one typically considers statistical query oracles with at best inverse polynomial accuracy, as in this regime the statistical query model provides a natural framework for studying the complexity of algorithms which always use sample access to a distribution to calculate expectation values of efficiently computable functions.\footnote{More specifically, in the regime of inverse polynomially accurate queries, i.e.,  $\tau = \Omega(1/\mathrm{poly}(n))$, \textit{any} statistical query algorithm (no matter its query complexity) yields a sample efficient algorithm in the sample model, as all queries to $\SQ_{\tau}(P)$ can be simulated using \textit{the same} set of samples from $\sample(P)$ \cite{diakonikolas2017statistical}. This is why lower bounds on the query complexity of an SQ algorithm do \textit{not} correspond to information-theoretic obstacles. They rather yield lower bounds on the \textit{computational complexity} of ``generic" algorithms in the sample model, i.e. those algorithms that simply simulate SQ access via access to the sample oracle.} We stress however that the opposite is not true, and that one \textit{cannot} simulate a sample oracle with a statistical query oracle, and therefore in principle it is possible that, for some computational problem, there exist sample efficient algorithms in the sample model but not in the statistical query model. 

Having fixed the different notions of access to a distribution that we will consider, we now define what it means to ``learn a distribution". In particular, as we have already mentioned, there are two distinct notions one could meaningfully consider. Informally, given some unknown target distribution $P$, we could ask that a learning algorithm, when given either SQ or sample oracle access to $P$, outputs an \emph{evaluator} for $P$ -- i.e. some function $\tilde{P}:\{0,1\}^n \rightarrow [0,1]$ which on input $x \in \{0,1\}^n$ outputs an estimate for $P(x)$, and therefore provides an approximate description of the distribution. This is perhaps the most intuitive notion of what it means to learn a probability distribution, and one which is often referred to as \textit{density modelling}, due to the fact that the evaluator allows one to model the probability density function of the unknown target distribution. However, in many practical settings one might not be interested in learning a full description of the probability distribution (an evaluator for the probability of events) but rather in being able to generate samples from the distribution. As such, instead of asking for an evaluator of the target distribution we could ask that the learning algorithm outputs a \emph{generator} for $P$ -- i.e. a probabilistic (quantum or classical) algorithm which when run generates samples from $P$. This task is often referred to as \textit{generative modelling}, due to the fact that the generator provides a model of the process via which samples from $P$ are generated. In order to formalize this, we start with the following definition of evaluators and generators.

\begin{definition}[Generators and evaluators]  Given some probability distribution $P\in\mathcal{D}_n$, we say that a classical (or quantum) algorithm $\gen_P$ is an efficient classical (quantum) generator for $P$ if $\gen_P$ produces samples in $\{0,1\}^n$ according to $P$, using $O(\mathrm{poly}(n))$ computational resources. 
In the case of a classical generator, we allow the algorithm to receive as input $m=O(\mathrm{poly}(n))$ uniformly random input bits. An algorithm $\eval_P:\{0,1\}^n\rightarrow [0,1]$ is an efficient evaluator for $P\in\mathcal{D}_n$ if for all $x\in\{0,1\}^n$ one has that $\eval_P(x) = P(x)$, and $\eval_P$ uses only $O(\mathrm{poly}(n))$ computational resources.
\end{definition}

Given the above definitions, we are now able to define formally the PAC framework for probabilistic modelling, which includes both generator and evaluator learning, and allows us to consider arbitrary models of oracle access to the unknown target distributions. Importantly, this framework also allows us to study the computational and statistical properties of probabilistic modelling algorithms, and to compare in a rigorous way quantum and classical learning algorithms. We start with the following definition of PAC generator and evaluator learners, which at a high level are learning algorithms which, when given oracle access to an unknown distribution, output with sufficiently high probability, a sufficiently accurate generator or evaluator.

\begin{definition}[PAC generator and evaluator learners]\label{d:PAC_learner}
An algorithm $\mathcal{A}$ is an $(\epsilon,\delta,O)$-PAC $\gen$-learner ($\eval$-learner) for $\mathcal{C}\subseteq \mathcal{D}_n$ if for all $P\in\mathcal{C}$, when given access to oracle $O(P)$, with probability at least $1-\delta$, $\mathcal{A}$ outputs a generator $\gen_Q$ (evaluator $\eval_Q$) for some $Q$ satisfying $\mathrm{d}_{\mathrm{TV}}(P,Q) \leq \epsilon$.
\end{definition}
Before proceeding, we re-iterate a few important aspects of the above definition (which are also discussed in detail in Ref.~\cite{Sweke2021quantumversus}). Firstly, we note that the learning algorithm in the above definition could be either quantum or classical. Indeed, one of the primary motivations of this work is to understand the potential advantages quantum learners could offer over classical learners for probabilistic modelling problems. Additionally, in the case of generator-learning the output generator could be either a quantum or classical generator, and we stress that one must be able to ``unplug" this generator from the oracle used during training -- i.e. the generator must be a completely independent algorithm for generating samples from the target distribution. For example, as mentioned in the introduction, QCBM based learning algorithms are a class of generative modelling algorithms whose output generator is a quantum circuit, which allows one to sample from the corresponding Born distribution by measuring the output state in the computational basis. Finally, we reiterate that while it is not perhaps a-priori clear why we would consider learning algorithms with access only to statistical queries, we note that many generic implicit generative modelling algorithms, both quantum and classical, when given access to a sample oracle use this access to approximate expectation values of functions (see Appendix~\ref{a:stat_query_algorithms}). As such, at least in the case of generative modelling, the statistical query model provides a natural framework for studying the complexity of learning problems with respect to known algorithms and methods. With this in hand, we can now define a variety of notions of \textit{efficient} PAC learnability of a distribution concept class.

\begin{definition}[Efficiently learnable distribution concept classes]\label{d:PAC_eff_learnable} Given a distribution concept class $\mathcal{C} \subseteq \mathcal{D}_n$ we define the randomized query complexity $\mathrm{RQC}_L(\mathcal{C},O,\delta,\epsilon,\gen)$ ($\mathrm{RQC}_L(\mathcal{C},O,\delta,\epsilon,\eval)$) as the smallest number of queries required by any $(\epsilon,\delta,O)$-PAC $\gen$-learner ($\eval$-learner) for $\mathcal{C}$. We say that $\mathcal{C}$ is sample-efficiently PAC $\gen$-learnable ($\eval$-learnable) with respect to oracle $O$ if for all $\epsilon,\delta \in (0,1)$
\begin{equation}
    \mathrm{RQC}_L(\mathcal{C},O,\delta,\epsilon,\gen (\eval)) = O\left(\mathrm{poly}\left(n,\frac{1}{\delta},\frac{1}{\epsilon}\right)\right).
\end{equation}
We say that a distribution concept class $\mathcal{C}$ is computationally-efficiently PAC $\gen$-learnable ($\eval$-learnable) with respect to oracle $O$ if it is sample-efficiently PAC $\gen$-learnable ($\eval$-learnable) with respect to oracle $O$, and in addition the sample-efficient learning algorithm also runs in time $\mathrm{O}(\mathrm{poly}(n,1/\delta,1/\epsilon))$ for all $\epsilon,\delta \in (0,1)$.
\end{definition}

\subsection{Local quantum circuit based distribution concept classes}\label{ss:lqc}

\begin{figure}
\centering
\includegraphics[width=0.7\linewidth]{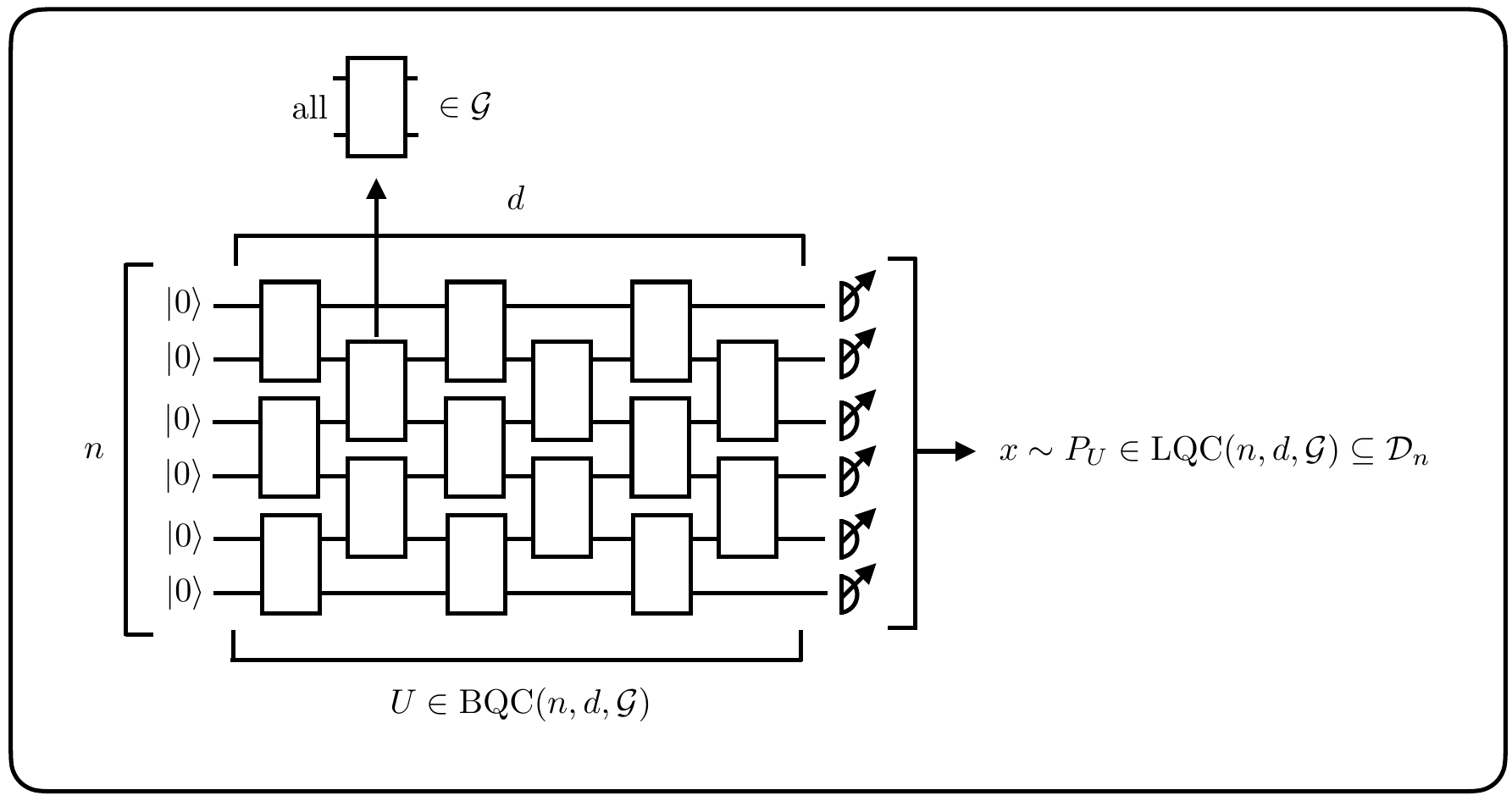}
\caption{An illustration of the brickwork quantum circuit architecture that we consider in this work (see also Definition \ref{d:bqc}). As per Definition \ref{d:concept_class}, we are concerned in this work with the learnability of the distributions obtained, as illustrated above, by measuring the output states of brickwork quantum circuits in the computational basis. 
}\label{fig:bqc}
\end{figure}

In this work we will be primarily concerned with the PAC learnability of the distributions obtained by measuring, in the computational basis, the output states of a specific class of local quantum circuits. In order to define this distribution concept class in a rigorous way, we start with the following definition of ``brickwork" quantum circuits, which is also illustrated in Figure \ref{fig:bqc}.

\begin{definition}[Brickwork quantum circuits with gate set $\mathcal{G}$]\label{d:bqc} Given some two-qubit gate-set $\mathcal{G}\subseteq \mathrm{U}(4)$ we denote by $\mathrm{BQC}(n,d,\mathcal{G})\subseteq \mathrm{U}(2^n)$ the set of unitaries which can be realized by a depth $d$ quantum circuit on $n$ qubits consisting only of nearest neighbour gates from $\mathcal{G}$. 
\end{definition}

While the above definition allows for an arbitrary two-qubit gate-set, we will predominantly be concerned with the two-qubit Clifford group, which we denote with $\mathrm{Cl}(4)$. In general we denote the $n$-qubit Clifford group as $\mathrm{Cl}(2^n)\subset \mathrm{U}(2^n)$. Given this definition, we proceed to define the classical probability distribution obtained by measuring the output state of a local quantum circuit in the computational basis. As the probabilities of events are derived from the amplitudes of the measured quantum state via the Born rule, we refer to this distribution as the Born distribution of the unitary which prepares the state.

\begin{definition}[Born distribution] Given some $n$-qubit unitary $U\in\mathrm{U}(2^n)$, we define the ``Born distribution" $P_U \in\mathcal{D}_n$ via
\begin{equation}
    P_U(x) = |\langle x|U|0^{\otimes n}\rangle|^2
\end{equation}
for all $x\in \{0,1\}^n$ - i.e. $P_U(x)$ is the probability of obtaining $x$ when measuring $U|0^{\otimes n}\rangle$ in the computational basis.
\end{definition}

With these definitions in hand, we can finally define the concept class of central interest to this work, namely the set of distributions obtained by measuring the output states of brickwork quantum circuits in the computational basis. Additionally, we will define the set of Born distributions corresponding to global Clifford unitaries, as we will later have reason to make use of this class of distributions.

\begin{definition}[Concept class $\mathrm{LQC}(n,d,\mathcal{G})$ and $\mathcal{D}_{\mathrm{Cl}(2^n)}$]\label{d:concept_class} Given some gate-set $\mathcal{G}\subseteq \mathrm{U}(4)$, for all $n,d$ we define ${\mathrm{LQC}(n,d,\mathcal{G})\subseteq\mathcal{D}_n}$ via
\begin{equation}
    \mathrm{LQC}(n,d,\mathcal{G}) = \left\{P_U \,|\, U\in \mathrm{BQC}(n,d,\mathcal{G})\right\}.
\end{equation}
Additionally, we define $\mathcal{D}_{\mathrm{Cl}(2^n)} \subset \mathcal{D}_n$ via
\begin{equation}
    \mathcal{D}_{\mathrm{Cl}(2^n)} = \{P_U\,|\,U\in\mathrm{Cl}(2^n)\}.
\end{equation}
\end{definition}
We note that $d_1 \leq d_2$ implies $\mathrm{LQC}(n,d_1,\mathcal{G}) \subseteq \mathrm{LQC}(n,d_2,\mathcal{G})$ and that $\mathrm{LQC}(n,d,\mathrm{Cl}(4)) \subseteq \mathcal{D}_{\mathrm{Cl}(2^n)}$ for all $d$. Finally, our proofs will often rely heavily on the fact that we can build a measure over the set of probability distributions $\mathrm{LQC}(n,d,\mathcal{G})$ by drawing gates of a circuit architecture uniformly at random, and then outputting the Born distribution of the global circuit unitary. In order to facilitate this, we define the following measure over $\mathrm{BQC}(n,D,\mathcal{G})$ which is induced by drawing gates in the circuit uniformly at random from the relevant gate set.

\begin{definition}[Induced measure over $\mathrm{BQC}(n,D,\mathcal{G})$]  We define $\mu(n,d,\mathcal{G})$ as the measure over $\mathrm{BQC}(n,d,\mathcal{G})$ which is induced by drawing gates from $\mathcal{G}$ uniformly. \label{d:induced_measure}
\end{definition}

\subsection{Decision problems in the statistical query model}\label{ss:SQ}

As mentioned in the very brief sketch of proof techniques given in the introduction, in order to obtain our first result we will rely heavily on a reduction between probabilistic modelling and a specific type of \textit{decision problem}, defined in Ref.~\cite{feldman2017general}, 
as follows:

\begin{definition}[$\SQ$ distribution-decision problem \cite{feldman2017general}] Given a set of distributions $\mathcal{C}\subset \mathcal{D}_n$, a reference distribution $D_0\in\mathcal{D}_n\setminus\mathcal{C}$, 
and some $(\delta,\tau) \in(0,1)$, we say that an algorithm $\mathcal{A}$ solves the distribution-decision problem $\mathrm{DEC}(D_0 \leftrightarrow\mathcal{C})$, with probability $1-\delta$, using oracle access to $\SQ_\tau$, if for all $P\in\mathcal{C}\cup \{D_0\}$
\begin{enumerate}
    \item when $P=D_0$ then $\mathrm{Pr}[\mathcal{A}^{\SQ_{\tau}(P)}\text{ outputs } 1]\geq 1-\delta$,
    \item when $P\in \mathcal{C}$ then $\mathrm{Pr}[\mathcal{A}^{\SQ_{\tau}(P)} \text{ outputs } 0]\geq 1-\delta$.
\end{enumerate}
We define the randomized query complexity $\mathrm{RQC}_D(D_0,\mathcal{C},\tau,\delta)$ as the smallest number of queries necessary for a randomized algorithm to solve the decision problem $\mathrm{DEC}(D_0 \leftrightarrow\mathcal{C})$, with probability $1-\delta$, using oracle access to $\SQ_\tau$.
\end{definition}
In order to gain some intuition for this type of decision problem we note that, given some $\tilde{\mathcal{C}}\subseteq\mathcal{D}_n$, when $\mathcal{C}=\tilde{\mathcal{C}}\setminus\mathcal{B}(D_0,\epsilon)$, then the decision problem $\mathrm{DEC}(D_0 \leftrightarrow\mathcal{C})$ is essentially equivalent to the problem of testing whether an unknown distribution is equal to the reference distribution $D_0$ or $\epsilon$ far from $D_0$, but with the additional promise that the unknown distribution is an element of the distribution concept class~$\tilde{\mathcal{C}}$. This observation allows us to use the standard property testing insight that learning is generically harder than testing to build a reduction between probabilistic modelling and a specific decision problem of the form just introduced. In particular, it is straightforward to show that ``learning implies deciding", i.e. that one can lower bound the randomized query complexity of learning $\mathcal{C}$ via the randomized query complexity of the decision problem $\mathrm{DEC}(D_0 \leftrightarrow\mathcal{C} \setminus\mathcal{B}(D_0,\epsilon))$. 

\begin{restatable}[Learning implies deciding]{lemma}{learningimpliesdeciding} \label{l:learning_to_decision}%
Assume $n\geq 1$,  $\epsilon,\delta,\tau\in(0,1)$, and $\mathcal{C}\subseteq\mathcal{D}_n$. Then, for all $D_0\in\mathcal{C}$ the following two inequalities hold
\begin{align} 
    \mathrm{RQC}_L(\mathcal{C},\mathrm{SQ}_{\tau},\delta,\epsilon,\gen) &\geq \mathrm{RQC}_D(D_0,\mathcal{C}\setminus\mathcal{B}(D_0,\epsilon),\tau,\delta),\\
    \mathrm{RQC}_L(\mathcal{C},\mathrm{SQ}_{\tau},\delta,\epsilon,\eval) &\geq \mathrm{RQC}_D(D_0,\mathcal{C}\setminus\mathcal{B}(D_0,\epsilon),\tau,\delta).
\end{align}
\end{restatable}
\begin{proof}
See Appendix \ref{app:l_to_d}.
\end{proof}

Our motivation for such a reduction comes from the fact that it allows us to exploit existing results from Feldman~\cite{feldman2017general}, which show that, in the statistical query model, the query complexity of a given decision problem is completely determined by the \textit{randomized statistical dimension} of the problem, which is defined as follows:

\begin{definition}[Randomized statistical dimension \cite{feldman2017general}]\label{d:RSD} Given some $\tau\in[0,1]$, we define the randomized statistical dimension $\mathrm{RSD}_{\tau}(D_0\leftrightarrow\mathcal{C})$ of the distribution-decision problem $\mathrm{DEC}(D_0 \leftrightarrow\mathcal{C})$ via
\begin{equation}\label{e:RSD_def}
    \mathrm{RSD}_{\tau}(D_0 \leftrightarrow\mathcal{C}) := \sup_{\nu\in S^{\mathcal{C}} }\left(\mathrm{frac}(\nu,D_0,\tau)^{-1}\right),
\end{equation}
where the supremum is over all probability measures $S^{\mathcal C}$ over the set $\mathcal{C}$, and $\mathrm{frac}(\nu,D_0,\tau)$ is defined via
\begin{equation}
    \mathrm{frac}(\nu,D_0,\tau) := \max_{\phi:\{0,1\}^n\rightarrow[-1,1]} \left\{\underset{D\sim\nu}{\pr}\left[ \left|D[\phi] - D_0[\phi]\right|> \tau\right] \right\}, \label{e:frac_feldman}
\end{equation}
where we have again used the shorthand notation $D[\phi] := \mathbb{E}_{x\sim D}[\phi(x)]$.
\end{definition}

We note that a statistical query via some function $\phi:\left\{ 0,1\right\} ^{n}\to\left[-1,1\right]$ allows one to distinguish, from the reference distribution $D_0$, all those distributions $D\in \mathcal{C}$ that satisfy $\left|D[\phi] - D_0[\phi]\right|>\tau$. 
Hence, we can think of each distinguishing function $\phi$ as covering a certain fraction of the class $\mathcal{C}$. In Ref. \cite{feldman2017general}, Feldman proved that the randomized statistical dimension as defined above equals precisely the size of a randomized cover on the whole class $\mathcal{C}$ by a measure over distinguishing functions $\phi:\left\{ 0,1\right\} ^{n}\to\left[-1,1\right]$. However, the utility of the randomized statistical dimension stems from the fact that Feldman was able to come up with a dual formulation in terms of a measure $\nu$ over the class of distributions $\mathcal{C}$, rather than over the distinguishing functions $\phi$. To illustrate this, note that the expression appearing in curly brackets in Eq. \eqref{e:frac_feldman}, 
\begin{equation}
    \left\{\underset{D\sim\nu}{\pr}\left[ \left|D[\phi] - D_0[\phi]\right|> \tau\right] \right\},
\end{equation}
is the probability that a distribution $D$, drawn randomly according to the measure $\nu$ on $\mathcal{C}$, can be distinguished from the reference distribution $D_0$ via a query to some fixed $\phi$.
For our purposes, it is helpful to point out the role of the measure $\nu$ when it comes to lower bounding the randomized statistical dimension: Due to the supremum in Eq.~\eqref{e:RSD_def}, any particular choice of measure $\nu$ leads to a value for $\mathrm{frac}(\nu,D_0,\tau)$, which in turn yields a lower bound on the randomized statistical dimension. However, to obtain the best possible bound, intuitively, we should choose the measure $\nu$ such that it is concentrated on distributions $D\in\mathcal{C}$ that are ``maximally hard" to distinguish from $D_0$. Such distributions will typically each require their own individual query $\phi$ in order to be distinguished from the reference distribution.

In light of this, we make particular use of the following lemma, which shows that the randomized query complexity of a decision problem can be lower bounded by the randomized statistical dimension.
\begin{lemma}[Randomized statistical dimension lower bounds randomized query complexity \cite{feldman2017general}]\label{l:RQC_from_RSD}
\begin{equation}
    \mathrm{RQC}_D(D_0,\mathcal{C},\tau,\delta)\geq \mathrm{RSD}_{\tau}(D_0 \leftrightarrow\mathcal{C})\cdot(1-2\delta).
\end{equation}
\end{lemma}
As such, by combining Lemma's \ref{l:learning_to_decision} and \ref{l:RQC_from_RSD} we see that in order to lower bound the query complexity of \textit{learning} a given distribution concept class $\mathcal{C}$, it is sufficient to lower bound the randomized statistical dimension of the decision problem $\mathrm{DEC}(D_0 \leftrightarrow\mathcal{C}\setminus\mathcal{B}(D_0,\epsilon))$. Additionally, we will also make use of the following observation that increasing the size of the set of distributions defining a decision problem can only increase the randomized statistical dimension, and therefore the randomized query complexity, of the associated decision problem. 

\begin{observation}[Randomized statistical dimension grows with the size of the concept class]\label{obs:RSD_only_grows}
Given two sets  of distributions $\mathcal{C}_1\subseteq \mathcal{C}_2 \subseteq\mathcal{D}_n$, a reference distribution $D_0\in\mathcal{D}_n\setminus\mathcal{C}_2$, and some $\tau\in(0,1)$, we have that
\begin{equation}
    \mathrm{RSD}_{\tau}(D_0 \leftrightarrow\mathcal{C}_2) \geq \mathrm{RSD}_{\tau}(D_0 \leftrightarrow\mathcal{C}_1).
\end{equation}
\end{observation}

\subsection{Linear algebra over
$\mathbb{F}_2^n$}

In order to prove Result \ref{r:informal_learnability} we exploit fundamental connections between the Born distributions of local Clifford circuits, and affine subspaces of the vector space $\mathbb{F}^n_2$. As such we review the required preliminaries here. In particular, we denote by $\mathbb{F}_2 = \{0,1\}$ the finite field of two elements. $\mathbb{F}^n_2$ is then the finite $n$-dimensional vector space over the field $\mathbb{F}_2$, whose elements are bit strings in $\{0,1\}^n$, equipped with entry wise addition modulo 2, which we denote with $\oplus$. We note that any $m$-dimensional subspace of $\mathbb{F}^n_2$ is isomorphic to $\mathbb{F}^m_2$, and can be described by a (non-unique) $n\times m$ binary matrix of full rank (containing basis vectors for the subspace). Additionally, we recall the following definition of an \textit{affine subspace}.

\begin{definition}[Affine subspace]
Let $V$ be a vector space over the field $\mathbb{F}$. A subset $M\subseteq V$ is called an affine subspace of $V$ if and only if there exists a vector $v\in V$ and a subspace $U\subseteq V$ such that
\begin{equation}
     M = v + U = \{v + u \,|\, u\in U\}.
\end{equation}
We define the dimension of $M$ via $\mathrm{dim}(M) = \mathrm{dim}(U)$.
\end{definition}

Given the above definition, we note that every $m$-dimensional affine subspace of $\mathbb{F}^n_2$ is fully specified by a (non-unique) tuple $(\mathbf{R},t)$, where $\mathbf{R}$ is an $n\times m$ full-rank binary matrix specifying the $m$-dimensional subspace $U\subseteq \mathbb{F}^n_2$, and $t\in\{0,1\}^n$ is the (non-unique) offset vector. More specifically, we say that such a tuple $(\mathbf{R},t)$ \textit{describes} an affine subspace $A\subseteq \mathbb{F}^n_2$ if 
\begin{equation}
    A = \{\mathbf{R}b \oplus t\,|\, b\in\mathbb{F}^m_2\}.
\end{equation}
Finally, given an $m$-dimensional affine subspace $A$ of $\mathbb{F}^n_2$, specified by the tuple $(\mathbf{R},t)$, we denote by $U_A \subseteq\mathcal{D}_n$ the uniform distribution over elements of $A$, i.e., 
the distribution for which for all $x\in\{0,1\}^n$
\begin{align}
    U_A(x) &= \begin{cases}\frac{1}{|A|} \text{ if } x\in A\\
    0 \text{ otherwise}
    \end{cases}\\
    &= \begin{cases}\frac{1}{2^m} \text{ if there exists } b\in\mathbb{F}^m_2 \text{ such that } x = \mathbf{R}b \oplus t\\
    0 \text{ otherwise.}
    \end{cases}
\end{align}

\section{Hardness of PAC learning the output distributions of local quantum circuits in the SQ model }\label{s:main_results}

We present in this section our first main result -- a formal version of Result \ref{r:informal_hardness} -- which is given below as Corollary~\ref{c:no_sample_efficient_generic}. In order to establish this result, we begin with the following theorem (whose proof is given in Section \ref{ss:proof_sketch}) which provides a lower bound on the randomized query complexity, in the statistical query model, of both generator-learning and evaluator-learning the output distributions of Clifford brickwork quantum circuits.

\begin{theorem}[Lower bound on the query complexity of learning local Clifford circuits in the SQ model]\label{t:RQCL_lower_bound}
For all $n$ large enough, all $d> 9$, and for all $\epsilon \in [0,1/6)$, 
\begin{align}
    \mathrm{RQC}_L(\mathrm{LQC}(n,d,\mathrm{Cl}(4)),\mathrm{SQ}_{\tau},\delta,\epsilon,\gen) &=\Omega\left(\tau^22^d(1-2\delta)\right) , \\
    \mathrm{RQC}_L(\mathrm{LQC}(n,d,\mathrm{Cl}(4)),\mathrm{SQ}_{\tau},\delta,\epsilon,\eval) &=\Omega\left(\tau^22^d(1-2\delta)\right).
\end{align}
\end{theorem}
We note that, perhaps surprisingly, the asymptotic query complexity lower bounds we obtain above are independent of the accuracy parameter $\epsilon$, provided it is suitably bounded. Additionally, we note that the lower bounds of Theorem~\ref{t:RQCL_lower_bound} depend on both the circuit depth $d$ and statistical query tolerance $\tau$. However, as mentioned and motivated in Section~\ref{ss:PAC_dist}, we are most naturally interested in the complexity of learning algorithms with respect to SQ oracles which provide expectation values of at best inverse polynomial accuracy $\tau$ -- i.e. in the setting where  $\tau =\Omega(1/\mathrm{poly}(n))$. In this setting, we have that super-logarithmic circuit depth $d = \omega(\log(n))$ is sufficient to obtain a super-polynomial lower bound on the query complexity of both PAC generator and evaluator learners for the output distributions of brickwork Clifford circuits. This then immediately implies the hardness, in the inverse polynomially accurate SQ model, of PAC learning the output distributions of super-logarithmically deep brickwork Clifford circuits. However, as increasing the size of a concept class can only increase the required query complexity, this also implies the hardness of PAC learning the output distributions of super-logarithmically deep brickwork circuits using \textit{any gate set which includes the Clifford group}. These observations are formalized in the following Corollary.

\begin{corollary}[Hardness of PAC learning local quantum circuits with inverse polynomially accurate statistical queries]\label{c:no_sample_efficient_generic}
Let $\mathcal{G}\subseteq\mathrm{U}(4)$ be any two-qubit gate-set satisfying $\mathrm{Cl}(4)\subseteq \mathcal{G}$. 
Then, for all $n$ large enough, for all $d=\omega(\log n)$, and for all $\tau = \Omega(1/\mathrm{poly}(n))$, the distribution concept class $\mathrm{LQC}(n,d,\mathcal{G})$ is not sample-efficiently PAC $\gen$-learnable or $\eval$-learnable with respect to the $\SQ_{\tau}$ oracle. 
\end{corollary}

\begin{proof}
Consider first the case $\mathcal{G}=\mathrm{Cl}(4)$. Using, Theorem \ref{t:RQCL_lower_bound} and taking $\tau$ and $d$ as per the statement of the Corollary immediately gives a super-polynomial lower bound (asymptotically with respect to $n$) for both $\mathrm{RQC}_L(\mathrm{LQC}(n,d,\mathrm{Cl}(4)),\mathrm{SQ}_{\tau},\delta,\epsilon,\gen)$ and $\mathrm{RQC}_L(\mathrm{LQC}(n,d,\mathrm{Cl}(4)),\mathrm{SQ}_{\tau},\delta,\epsilon,\eval)$, which implies the statement of the Corollary for $\mathcal{G}=\mathrm{Cl}(4)$. The case $\mathcal{G} \supseteq \mathrm{Cl}(4)$ then follows from the observation that, for any two distribution concept classes $\mathcal{C}_1$ and $\mathcal{C}_2$ satisfying $\mathcal{C}_1\subseteq\mathcal{C}_2$, both a $\gen$ or $\eval$ learner for $\mathcal{C}_2$ is immediately a learner for $\mathcal{C}_1$, and therefore 
\begin{align}
    \mathrm{RQC}_L(\mathcal{C}_2,\mathrm{SQ}_{\tau},\delta,\epsilon,\gen) &\geq \mathrm{RQC}_L(\mathcal{C}_1,\mathrm{SQ}_{\tau},\delta,\epsilon,\gen) , \\
    \mathrm{RQC}_L(\mathcal{C}_2,\mathrm{SQ}_{\tau},\delta,\epsilon,\eval) &\geq \mathrm{RQC}_L(\mathcal{C}_1,\mathrm{SQ}_{\tau},\delta,\epsilon,\eval).
\end{align}
\end{proof}

\begin{figure}[h!]
\centering
\includegraphics[width=0.4\linewidth]{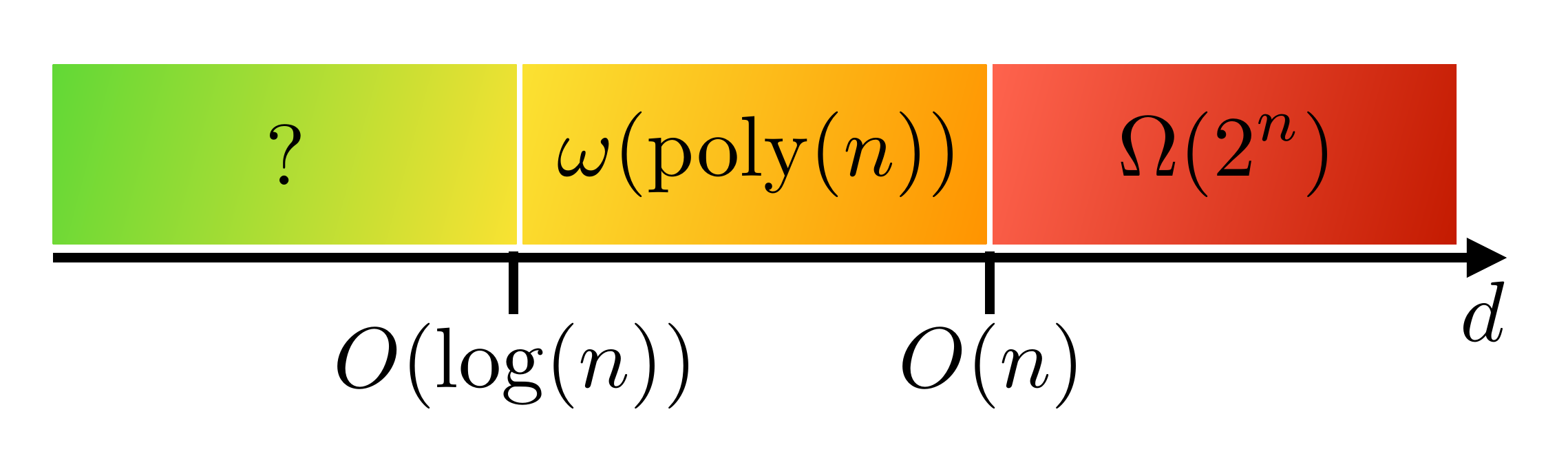}
\caption{Scaling of the query complexity lower bound in Theorem \ref{t:RQCL_lower_bound}, with respect to circuit depth $d$, in the regime where $\tau =\Omega(1/\mathrm{poly}(n))$. As described in Section \ref{ss:proof_sketch}, in order to prove Theorem \ref{t:RQCL_lower_bound}, we first restrict our attention to the case of linear depth circuits, i.e. to the right-hand side of the figure. More specifically, in Section \ref{ss:linear_depth}, we show how to obtain the exponential query complexity lower bound in this regime. Given this, in Section \ref{ss:sub_linear_depth}, we extend our results to the regime of sub-linear depth, i.e., we obtain the super-polynomial query complexity lower bound indicated in the middle part of the figure. Lastly, we note that our lower bound is polynomial in the regime of $d=O(\log(n))$. Hence, we leave open the question of SQ hardness in this regime since our bound might not be tight.}\label{fig:complexity_phase_diagram}
\end{figure}

Let us stress that, as Corollary \ref{c:no_sample_efficient_generic} is concerned with \textit{query complexity}, it applies to both \textit{classical and quantum} learning algorithms which use statistical queries. As discussed in Appendix \ref{a:stat_query_algorithms}, many generic generative modelling algorithms of practical interest can be efficiently simulated in the SQ model, and are therefore under the domain of applicability of this result. As such, Corollary \ref{c:no_sample_efficient_generic} strongly limits the potential for using the output distributions of local quantum circuits to provide a separation between the power of quantum and classical generative modelling algorithms. Additionally, we mention again that the concept class of super-logarithmically deep nearest neighbour Clifford circuits is \textit{classically simulatable} \cite{gottesman1998heisenberg}. As such, Corollary \ref{c:no_sample_efficient_generic} establishes that learning a generator or an evaluator from statistical queries can be hard, for both quantum and classical learning algorithms, even when outputting a classical generator from a circuit description can be done efficiently! Finally, as has been noted in the proof of Corollary \ref{c:no_sample_efficient_generic}, for inverse polynomially accurate SQ queries, super-logarithmic circuit depth is enough to ensure a super-polynomial lower bound on the randomized query complexity - which then implies the non-existence of efficient learning algorithms. However we note, as illustrated in Fig. \ref{fig:complexity_phase_diagram}, that that combination of inverse polynomially accurate SQ queries and polynomial depth circuits would give rise to an exponential lower bound on the query complexity. 

\begin{remark}[Generalization to universal gate sets] \label{re:universal_gate_sets}
Corollary \ref{c:no_sample_efficient_generic} establishes the hardness of SQ learning the Born distributions of local quantum circuits that use gates from any two-qubit gate set $\mathcal{G}$ that contains the two-qubit Clifford group $\mathrm{Cl}(4)$. A natural follow-up question is whether similar hardness results can be obtained for alternative gate sets not containing the Clifford group. Many interesting such gate sets exist. In fact, it is known that \textit{any} entangling two-qubit gate, together with arbitrary single-qubit gates, is universal \cite{brylinski2002universal, bremner2002practical}. Here, we remark that the proof techniques we use to establish the query complexity lower bounds given in Theorem \ref{t:RQCL_lower_bound} are indeed sufficiently general to be adapted to any universal gate set $\mathcal{G}$. This is because our proof of Theorem \ref{t:RQCL_lower_bound} does not rely on the algebraic properties of Clifford circuits and their Born distributions. Rather, it relies on the fact that the Clifford group is sufficiently evenly distributed over the unitary group. More specifically, we use that it forms a unitary 2-design (see Appendix \ref{app:moments} for a definition). Additionally, we use that any global Clifford unitary $U\in \mathrm{Cl}(2^n)$ can be implemented via a nearest-neighbor Clifford circuit in linear depth~\cite{Bravyi_2021}. It is known that local random quantum circuits with gates drawn from any universal gate set also converge (at least approximately) to a unitary $t$-design in linear depth, more precisely at depth $d=O(n\mathrm{poly}(t))$ \cite{brandaoLocalRandomQuantum2016,harrow2018approximate,haferkampImprovedSpectralGaps2021}. Given this, and using higher moments, our proof techniques can be adapted to local quantum circuits based on any universal gate set. Here, we choose to restrict the presentation to gate sets $\mathcal{G}$ satisfying $\mathrm{Cl}(4)\subseteq\mathcal{G}$ for reasons of brevity and clarity.
\end{remark}

\section{Proof of Theorem \ref{t:RQCL_lower_bound}}\label{ss:proof_sketch}

\begin{figure}[h!]
\centering
\includegraphics[width=1\linewidth]{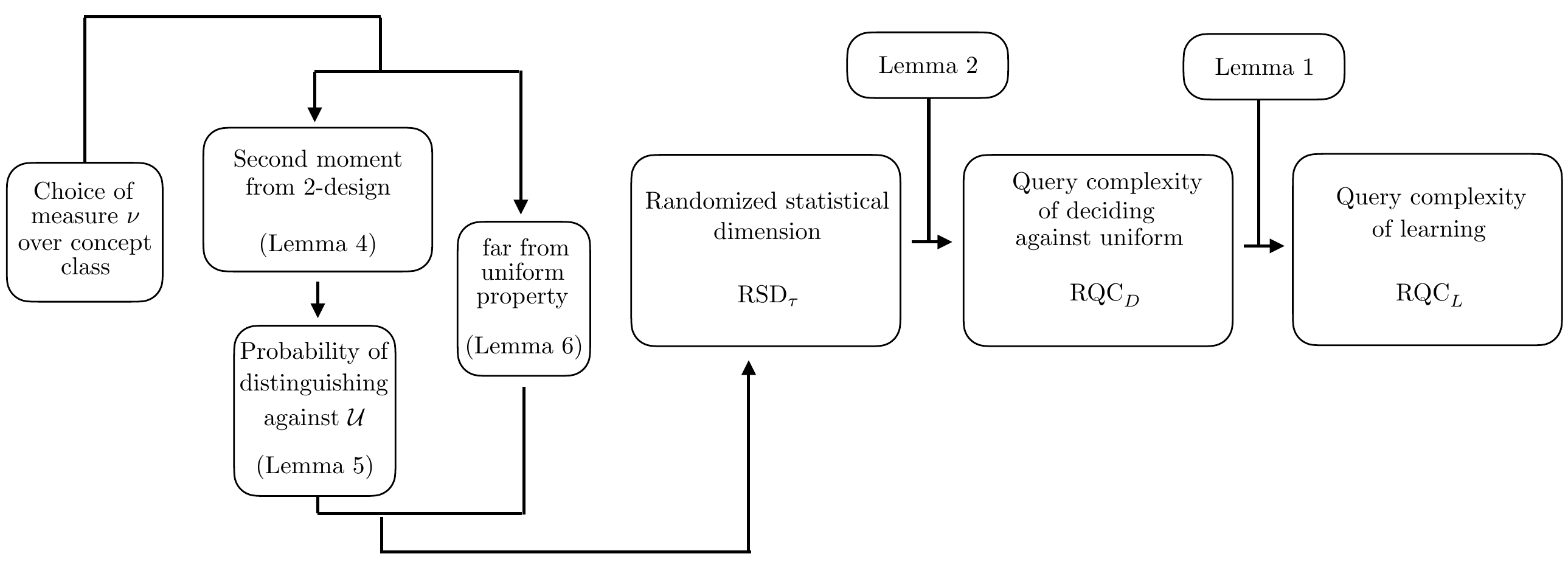}
\caption{Overview of the strategy used for lower bounding the statistical query complexity of learning the output distributions of local quantum circuits. On  the right hand side of the figure we illustrate the fact that lower bounds on the randomized statistical dimension of a suitable decision problem, give rise -- via Lemmas \ref{l:learning_to_decision} and \ref{l:RQC_from_RSD} -- to query complexity lower bounds for learning. On the left hand side of the figure we illustrate the ingredients used, in Section \ref{ss:linear_depth}, to lower bound the randomized statistical dimension of a suitable decision problem.}\label{fig:proof_sketch}
\end{figure}

We provide in this section a proof for the lower bounds on
\begin{align}
\mathrm{RQC}_L(\mathrm{LQC}(n,d,\mathrm{Cl}(4)),\mathrm{SQ}_{\tau},\delta,\epsilon,\gen), \\ \mathrm{RQC}_L(\mathrm{LQC}(n,d,\mathrm{Cl}(4)),\mathrm{SQ}_{\tau},\delta,\epsilon,\eval),
\end{align}
given in Theorem \ref{t:RQCL_lower_bound}. As discussed in Section \ref{ss:SQ}, these query complexity lower bounds can be obtained by proving a lower bound on the randomized statistical dimension of a suitably constructed decision problem
$$\mathrm{DEC}(D_0 \leftrightarrow\mathrm{LQC}(n,d,\mathrm{Cl}(4))\setminus\mathcal{B}(D_0,\epsilon)).$$
The above decision problem singles out one Born distribution $D_0\in \mathrm{LQC}(n,d,\mathrm{Cl}(4))$ as a reference distribution. The task is then to decide, given SQ oracle access to an unknown distribution $P\in\mathrm{LQC}(n,d,\mathrm{Cl}(4)) $, whether $P=D_0$ or is at least $\epsilon$ far from $D_0$ in total variation distance\footnote{It is important to note that this decision problem is close, but not quite the same, as the verification problem studied in Ref. \cite{hangleiter2019sample}. In particular, here we have \textit{a promise} that the unknown distribution $P$ is an element of the concept class $\mathrm{LQC}(n,d,\mathrm{Cl}(4))$. This promise is essential to the reduction between learning and deciding, but precludes the use of optimality results from identity testing \cite{
valiant2017automatic}, in which there is no such promise.}.
We note that, as a result of the reduction from deciding to learning given in Lemma \ref{l:learning_to_decision} of Section \ref{ss:SQ}, we are free to choose whichever reference distribution $D_0\in\mathrm{LQC}(n,d,\mathrm{Cl}(4))$  allows us to obtain the tightest lower bound.
We hence aim to choose $D_0$ such that it is ``maximally hard" to distinguish from the rest of the concept class via statistical queries. Intuitively, this will be the case whenever the concept class contains many distributions that cannot be distinguished from $D_0$ via the same statistical queries.

A natural candidate for a good reference distribution $D_0$ is the uniform distribution $\mathcal{U}$. Intuitively, this is because our concept class contains a large number of ``flat" distributions that will each require their own statistical query in order to be distinguished from the uniform distribution. Indeed, it is known that the Born distributions of local quantum circuits are typically exponentially flat at sufficient circuit depth \cite{hangleiter2019sample}.
Additionally, we note that as the uniform distribution can be straightforwardly generated by measuring the output state of a Clifford circuit of unit depth, we indeed have that $\mathcal{U}\in\mathrm{LQC}(n,d,\mathrm{Cl}(4))$ for all $d\geq 1$.
Given this intuition, we will therefore proceed by lower bounding the randomized statistical dimension (RSD) of the decision problem $\mathrm{DEC}(\mathcal{U} \leftrightarrow\mathrm{LQC}(n,d,\mathrm{Cl}(4))\setminus\mathcal{B}(\mathcal{U},\epsilon))$, which we denote as $$ \mathrm{RSD}_{\tau}(\mathcal{U}\leftrightarrow\mathrm{LQC}(n,d,\mathrm{Cl}(4))\setminus\mathcal{B}(\mathcal{U},\epsilon)).$$
While one would naturally expect the RSD to depend on the accuracy of the statistical queries ${\tau}$, the circuit depth $d$, number of qubits $n$ and accuracy $\epsilon$, we will find an aymptotic lower bound which, provided $\epsilon$ is suitably bounded, depends only on $n,\tau$ and $d$. As mentioned before, we are naturally most interested in setting $\tau =\Omega(1/\mathrm{poly}(n))$, because this accuracy corresponds to the regime in which one can simulate statistical queries via polynomially many samples. With $\tau$ ``fixed'' to $\tau =\Omega(1/\mathrm{poly}(n))$, our primary goal is then to determine the smallest depth $d$ giving rise to super-polynomial (in $n$) query complexity lower bounds, and hence to hardness of SQ learning.  While Theorem~\ref{t:RQCL_lower_bound} establishes this hardness for super-logarithmic depth circuits, we will prove this result in two steps:
\begin{enumerate}
\item \textbf{Linear depth:} As a warm-up, in the first part, we consider output distributions of nearest-neighbor Clifford circuits of linear depth. In particular, we will choose $d\geq 9n$ as it is known that this depth is sufficient to implement any Clifford unitary $U\in \mathrm{Cl}(2^n)$ exactly in a nearest-neighbor circuit architecture \citep{Bravyi_2021}. In our notation, this implies that, for all $d\geq 9n$, we have that 
\begin{equation*}
	\mathrm{BQC}(n,d,\mathrm{Cl}(4)) = \mathrm{Cl}(2^n),
\end{equation*}
and therefore that $\mathrm{LQC}(n,d,\mathrm{Cl}(4)) = \mathcal{D}_{\mathrm{Cl}(2^n)}$.
Exploiting this completeness property of the concept class, along with properties of the Born distributions of global Clifford unitaries, we are able to prove a lower bound on the randomized statistical dimension $\mathrm{RSD}_{\tau}(\mathcal{U}\leftrightarrow\mathrm{LQC}(n,d,\mathrm{Cl}(4))\setminus\mathcal{B}(\mathcal{U},\epsilon))$ which grows \textit{exponentially} in $n$, whenever $\tau =\Omega(1/\mathrm{poly}(n))$.

\item \textbf{Extension to sublinear depth:} The \textit{exponential} lower bound from the linear-depth case suggests that one might be still be able to achieve a \textit{super-polynomial} lower bound at sub-linear circuit depths. In the second part of the proof, we will show that this is indeed the case. In fact, we demonstrate that we are able to trade off circuit depth against query complexity. The technical difficulty arises from the fact that when we reduce the depth to $d<n$, then some global Clifford unitaries cannot be implemented anymore so that
\begin{equation*}
 \mathrm{BQC}(n,d,\mathrm{Cl}(4))\subset \mathrm{Cl}(2^n).
\end{equation*} 
Consequently, also the concept class $\mathrm{LQC}(n,d,\mathrm{Cl}(4))$ containing the corresponding Born distribution of the circuits gets smaller. Characterizing precisely which Clifford Born distributions drop out and which are still present at a certain depth seems difficult. Instead, building on the insights from the linear-depth case, we will show two different approaches for extending our bounds to sub-linear depth that get around this difficulty. Both approaches let us establish a trade-off between the depth $d$ -- and hence the size of the concept class -- and the number of required statistical queries.
\end{enumerate}

\subsection{Warm up: Linear circuit depth}
\label{ss:linear_depth}
\noindent In this section, we will prove the following lemma, which -- as illustrated in Figure \ref{fig:proof_sketch} -- after applying the reductions from Lemmas \ref{l:learning_to_decision} and \ref{l:RQC_from_RSD}, gives rise to a version of Theorem \ref{t:RQCL_lower_bound}, restricted to the setting of linear depth local quantum circuits . 

\begin{lemma}[Restriction of Theorem \ref{t:RQCL_lower_bound} for linear depth quantum circuits]\label{l:RSD_linear_depth}
For all $n$ large enough, all $d\geq 9n$ and all $\epsilon \in [0,1/6)$ it holds that
\begin{equation}
    \mathrm{RSD}_{\tau}(\mathcal{U}\leftrightarrow\mathrm{LQC}(n,d,\mathrm{Cl}(4))\setminus \mathcal{B}(\mathcal{U},\epsilon)) =  \mathrm{RSD}_{\tau}(\mathcal{U}\leftrightarrow\mathcal{D}_{\mathrm{Cl}(2^n)}\setminus \mathcal{B}(\mathcal{U},\epsilon))= \Omega\left(\tau^2 2^n\right).
\end{equation}
\end{lemma}

As stressed before, this immediately implies, for all $d\geq 9n$, statistical query complexity lower bounds for learning $\mathrm{LQC}(n,d,\mathrm{Cl}(4))$ which are exponential in $n$, whenever $\tau =\Omega(1/\mathrm{poly}(n))$.
This choice of depth is deliberate, since it follows from Ref. \cite{Bravyi_2021} that by using a nearest-neighbour Clifford circuit of depth at most $d=9n$ one can implement any global Clifford unitary $U\in\mathrm{Cl}(2^n)$.
Hence, we have that, for all $d\geq 9n$,
\begin{equation}
\mathrm{LQC}(n,d,\mathrm{Cl}(4)) = \mathcal{D}_{\mathrm{Cl}(2^n)}.
\end{equation}
Thus, when proving Lemma \ref{l:RSD_linear_depth}, the decision problem of interest is with respect to the Born distributions of the global $n$-qubit Clifford group
\begin{equation}
\mathrm{DEC}(\mathcal{U} \leftrightarrow\mathrm{LQC}(n,d,\mathrm{Cl}(4))\setminus\mathcal{B}(\mathcal{U},\epsilon)) = \mathrm{DEC}(\mathcal{U} \leftrightarrow\mathcal{D}_{\mathrm{Cl}(2^n)}\setminus\mathcal{B}(\mathcal{U},\epsilon))  .
\label{e:global_Clifford_decision}
\end{equation}
From Definition \ref{d:RSD}, it follows that the RSD of this decision problem is lower bounded by
\begin{equation}
    \mathrm{RSD}_{\tau}(\mathcal{U}\leftrightarrow \mathrm{Cl}(2^n)\setminus\mathcal{B}(\mathcal{U},\epsilon)) \geq \left(\max_{\phi:\{0,1\}^n\rightarrow[-1,1]} \left\{\underset{P\sim\nu}{\pr}\left[\left|P[\phi] - D_0[\phi]\right|> \tau\right]\right \}\right)^{-1}\label{e:RSD_global_Cliffords}
\end{equation}
for any measure $\nu$ over $\mathcal{D}_{\mathrm{Cl}(2^n)}\setminus\mathcal{B}(\mathcal{U},\epsilon)$.  This is because the RSD is actually defined as the supremum of the RHS of Eq. \eqref{e:RSD_global_Cliffords} over all possible measures $\nu$ over $\mathcal{D}_{\mathrm{Cl}(2^n)}\setminus\mathcal{B}(\mathcal{U},\epsilon)$.
When proving Lemma \ref{l:RSD_linear_depth}, we will take $\nu$ to be the measure induced by drawing a uniformly random global Clifford $U\sim \mathrm{Cl}(2^n)$  and post-selecting on its Born distribution $P_U$ 
being at least $\epsilon$ far from uniform in total variation distance. That is, $\nu$ is can be defined via the following procedure for sampling from $\nu$:
\begin{enumerate}
    \item Draw $U\sim \mathrm{Cl}(2^n)$.
    \begin{enumerate}
        \item If $\mathrm{d}_{\mathrm{TV}}(P_U,\mathcal{U}) > \epsilon$, output $P_U$.
        \item Else, if $P_U\in\mathcal{B}(\mathcal{U},\epsilon)$, resample from the uniform measure over $\mathrm{Cl}(2^n)$.
    \end{enumerate}
\end{enumerate}
It follows from the definition of the conditional probability $P(A|B)=P(A\cap B)/P(B)$ that, for all $\phi:\{0,1\}^n\rightarrow[-1,1]$,
\begin{equation}
\label{e:probability_fraction_global_Clifford}
    \underset{P\sim\nu}{\pr}\left[\left|P[\phi] - \mathcal{U}[\phi]\right|> \tau\right] 
    \leq \frac{\underset{U\sim\mathrm{Cl}(2^n)}{\pr}\left[ \left|P_U[\phi] - \mathcal{U}[\phi]\right|> \tau \right]}{\underset{U\sim\mathrm{Cl}(2^n)}{\pr}\left[ \mathrm{d}_{\mathrm{TV}}(P_U,\mathcal{U}) > \epsilon \right]}.
\end{equation}
Our goal is to upper bound the fraction appearing on the right hand side of Eq. \eqref{e:probability_fraction_global_Clifford}.
We will do so by finding bounds for the denominator and the numerator separately. Note that, due to our choice of $\nu$, the fraction involves probabilities over unitaries drawn uniformly at random from the Clifford group $\mathrm{Cl}(2^n)$. It turns out that expectation values of the form $\underset{U\sim \mathrm{Cl}(2^n)}{\mathbb{E}}f(U,\overline{U})$ can be evaluated exactly analytically as long as $f$ is a polynomial of degree at most $\deg(f)\leq3$ in the entries of $U$ and its complex conjugate $\overline U$. This will allows us to greatly simplify our computations when bounding the numerator. More specifically, we will make use of the following expressions for the first and second moment of output probabilities $P_U(x)$. 

\begin{restatable}[Clifford moments]{lemma}{cliffordmoments} \label{l:Clifford_moments}
\begin{align}
    	\underset{U\sim \mathrm{Cl}(2^n)}{\mathbb{E}} \left[ P_U(x) \right] 
	&=\frac{1}{2^n} \label{eq:Clifford_first_mom},\\
	\underset{U\sim \mathrm{Cl}(2^n)}{\mathbb{E}} \left[ P_U(x) P_U(y) \right] 
	& =\frac{1}{2^n(2^n+1)}[1 + \delta_{x,y}].
	\label{eq:Clifford_second_mom}
\end{align}
\end{restatable}

\begin{proof}
See Appendix \ref{app:moments}.
\end{proof}
Given these expressions, we can now start to bound the terms appearing in 
Eq.~\eqref{e:probability_fraction_global_Clifford}. Let us start with the numerator. We prove the following upper bound.

\begin{lemma}[Probability of distinguishing from $\mathcal{U}$ -- numerator of Eq. \eqref{e:probability_fraction_global_Clifford}]\label{l:numerator_linear_depth}
Assume $n$ large enough and $d\geq 9n$. Then for all $\phi:\{0,1\}^n\rightarrow [-1,1]$ one has that
\begin{equation}
    \underset{U\sim \mathrm{Cl}(2^n)}{\pr}\left[ \left|P_U[\phi] - \mathcal{U}[\phi]\right|> \tau \right] = O\left(\frac{1}{2^n\tau^2}  \right).
\end{equation}
\end{lemma}

\begin{proof}
Using the first moment from Eq. \eqref{eq:Clifford_first_mom}, we find that 
\begin{equation}
 	\underset{U\sim \mathrm{Cl}(2^n)}{\mathbb{E}} \left[ P_U[\phi]\right] 
 	= \sum_{x\in\{0,1\}^n} \left[\underset{U\sim \mathrm{Cl}(2^n)}{\mathbb{E}} \left[ P_U(x) \right] \phi(x)\right] =\mathcal{U}[\phi].
\end{equation}
By Chebyshev's inequality, for any $\tau>0$
\begin{equation}
    \underset{U\sim\mathrm{Cl}(2^n)}{\pr}\left[ \left|P_U[\phi] - \mathcal{U}[\phi]\right|> \tau \right] \leq \frac{\mathrm{Var}\left[P_U[\phi]\right]}{\tau^{2}}.
\end{equation}
The variance is given by
\begin{align}
	\mathrm{Var}\left[P_U[\phi]\right]
	&=\underset{U\sim\mathrm{Cl}(2^n)}{\mathbb{E}} \left[ P_U[\phi]^2 \right] - \left[\underset{U\sim\mathrm{Cl}(2^n)}{\mathbb{E}} \left[ P_U[\phi]\right]\right]^2 \\
	&= \sum_{x}\sum_{y}\phi(x)\phi(y) \left[\underset{U\sim\mathrm{Cl}(2^n)}{\mathbb{E}} \left[ P_U(x) P_U(y) \right]-\frac{1}{2^{2n}}\right].
\end{align}
Inserting the second moment from Eq. \eqref{eq:Clifford_second_mom} and bounding $\phi(x)\phi(y)\leq 1$, we find
\begin{align}
	\mathrm{Var}\left[P_U[\phi]\right]
	&= \sum_{x}\sum_{y}\phi(x)\phi(y) \left[ \frac{1}{2^n(2^n+1)}[1 + \delta_{x,y}] -\frac{1}{2^{2n}}\right]\\
	&\leq \frac{1}{2^n}
\end{align}
from which the claim follows.
\end{proof}
For the denominator, we show the following lower bound:
\begin{lemma}[Global random Clifford output distributions are far from uniform]
\label{l:pr_outside_ball_linear_depth}
Assume $n,d\geq 2$. Then for any $\epsilon \in [0,1/6]$,
\begin{equation}
    \underset{U\sim\mathrm{Cl}(2^n)}{\pr}\left[\mathrm{d}_{\mathrm{TV}}(P_U,\mathcal{U}) \geq \epsilon \right] \geq \frac{1/6-\epsilon}{1-\epsilon}.
\end{equation}
\end{lemma}
\begin{proof}
See Appendix \ref{app:pr_outside_ball}.
\end{proof}
Finally, with all the pieces in place, the proof of Lemma \ref{l:RSD_linear_depth} is straightforward. One simply substitutes the expressions from Lemma \ref{l:numerator_linear_depth} and Lemma \ref{l:pr_outside_ball_linear_depth} into Eq. \eqref{e:probability_fraction_global_Clifford}. As illustrated in Fig. \ref{fig:proof_sketch}, one then obtains a restricted version of Theorem~\ref{t:RQCL_lower_bound} by first using the relationship between the randomized statistical dimension and the randomized query complexity of decision problems (Lemma \ref{l:RQC_from_RSD}), and then applying the reduction between learning and deciding (Lemma \ref{l:learning_to_decision}).

Additionally, following up on Remark \ref{re:universal_gate_sets}, we point out that the crucial ingredient to prove Lemma \ref{l:numerator_linear_depth} is the 2-design property of the Clifford group, which leads to the moments given in Lemma \ref{l:Clifford_moments}. Hence, an analogous version of this Lemma could be derived for random circuits based on any universal gate set, since such circuits converge to unitary $t$-designs in depth $d=O(n)$, for any constant $t$. Furthermore, an analogous version of Lemma \ref{l:pr_outside_ball_linear_depth} can be derived whenever the underlying circuit ensemble forms at least an approximate unitary 4-design.

\subsection{Extension to sub-linear circuit depth}
\label{ss:sub_linear_depth}

In the previous section we established a lower bound for the randomized statistical dimension of a suitable decision problem, which leads to a restriction of Theorem \ref{t:RQCL_lower_bound} which holds only for linear depth circuits. In this section we provide two different techniques for proving Theorem \ref{t:RQCL_lower_bound} by extending these lower bounds to the case of sub-linear depth circuits, at the cost of decreased query complexity. As mentioned before, the primary difficulty we tackle here is the fact that for sub-linear circuit depth $d<n$ the concept class $\mathrm{LQC}(n,d,\mathrm{Cl}(4))$ is a \textit{strict} subset of the set of global Clifford Born distributions, i.e.,
$$\mathrm{LQC}(n,d,\mathrm{Cl}(4))\subset\mathcal{D}_{\mathrm{Cl}(2^n)},$$
and we therefore cannot rely straightforwardly on properties of the global Clifford unitaries. 
While the first approach detailed below provides a slightly weaker (but still super-polynomial) query complexity lower bound than that stated in Theorem \ref{t:RQCL_lower_bound}, we provide two alternative approaches due to the fact that extensions of either of these techniques may facilitate progress on the open questions and conjectures listed in Section \ref{s:discussion}.

\subsubsection{First approach: sub-linear circuit depth Clifford moments via random circuit techniques}\label{sss:random_circuits}

The first approach follows essentially the same steps as for the linear depth case, i.e. we aim to obtain a lower bound on the RSD of the decision problem 
$$\mathrm{DEC}(\mathcal{U} \leftrightarrow\mathrm{LQC}(n,d,\mathrm{Cl}(4))\setminus\mathcal{B}(\mathcal{U},\epsilon)),$$ which is applicable even in the case when $d<9n$.
The only difference arises from the issue that, in the case when $d<9n$,
we need to come up with a different measure $\nu$ that is only supported on elements that are still present in the concept class $\mathrm{LQC}(n,d,\mathrm{Cl}(4))$.
A natural choice for $\nu$ is the measure $\mu(n,d,\mathrm{Cl}(4))$ introduced in Definition \ref{d:induced_measure}. Essentially, instead of drawing a global Clifford unitary as considered in the linear-depth case, we now draw a random Clifford circuit of depth $d$ by drawing the individual gates in the circuit architecture independently from the 2-qubit Clifford group $\mathrm{Cl}(4)$.
To be precise, we will take $\nu$ to be the measure over the set $\mathrm{LQC}(n,d,\mathrm{Cl}(4))\setminus\mathcal{B}(\mathcal{U},\epsilon)$ defined by the following procedure for sampling from $\nu$:
\begin{enumerate}
    \item Draw $U\sim \mu(n,d,\mathrm{Cl}(4))$
    \begin{enumerate}
        \item If $\mathrm{d}_{\mathrm{TV}}(P_U,\mathcal{U}) > \epsilon$, output $P_U$.
        \item Else, if $P_U\in\mathcal{B}(\mathcal{U}, \epsilon)$, reject and resample from $\mu(n,d,\mathrm{Cl}(4))$.
    \end{enumerate}
\end{enumerate}
As in the linear-depth case, we now have that
\begin{equation}
    \underset{P\sim\nu}{\pr}\left[\left|P[\phi] - \mathcal{U}[\phi]\right|> \tau\right] 
    \leq \frac{\underset{U\sim\mu(n,d,\mathrm{Cl}(4))}{\pr}\left[ \left|P_U[\phi] - \mathcal{U}[\phi]\right|> \tau \right]}{\underset{U\sim\mu(n,d,\mathrm{Cl}(4))}{\pr}\left[ \mathrm{d}_{\mathrm{TV}}(P_U,\mathcal{U}) > \epsilon \right]},
    \label{e:probability_fraction_random_circuit}
\end{equation}
where in contrast to Eq. \eqref{e:probability_fraction_global_Clifford}, the probabilities in the fraction on the RHS of Eq. \eqref{e:probability_fraction_random_circuit} are now with respect to a randomly drawn nearest-neighbor Clifford circuit rather than a randomly drawn global Clifford unitary. Luckily, it turns out that moments with respect to random nearest-neighbor Clifford circuits, i.e. moments with respect to $\mu(n,d,\mathrm{Cl}(4))$, can be bounded using techniques from the existing literature on random circuits. Specifically, slightly modifying results on the collision probability of random quantum circuits~\cite{barak2021spoofing,dalzell2020random} allows us to compute the following first and second moment bounds, from which bounds on the numerator and denominator follow as per the linear depth case. In particular, the following lemma is adapted from Section 6.3 of Ref.~\cite{barak2021spoofing}.

\begin{restatable}[Restricted depth random nearest-neighbor Clifford circuit moments -- adapted from Ref.~\cite{barak2021spoofing}]{lemma}{restricteddepthcliffordmoments} \label{l:brickwork_moments}%
\begin{align}
    	\underset{U\sim \mu(n,d,\mathrm{Cl}(4))}{\mathbb{E}} \left[ P_U(x) \right] 
	&= \frac{1}{2^n} \label{eq:brickwork_first_mom_2} ,\\
\underset{U\sim \mu(n,d,\mathrm{Cl}(4))}{\mathbb{E}} \left[ P_U(x) P_U(y) \right] 
	&\leq 
	\frac{1}{2^{2n}}\left[
	(1+\delta_{x,y})\left[1+n\left(\frac{4}{5}\right)^{d}\right]\right].\label{eq:brickwork_second_mom}
\end{align}
\end{restatable}

\begin{proof}
See Appendix \ref{app:moments}.
\end{proof}

Using these expressions, bounding the randomized statistical dimension proceeds  completely analogously to the linear-depth case. In particular, we can directly prove the following bound on the numerator of Eq. \eqref{e:probability_fraction_random_circuit}. 
\begin{lemma}[Probability of distinguishing from $\mathcal{U}$ -- numerator of Eq. \eqref{e:probability_fraction_random_circuit}]
\label{l:numerator_restricted_depth}
Assume $n$ large enough and $d=\Omega(\log(n))$. Then for all $\phi:\{0,1\}^n\rightarrow [-1,1]$ one has that
\begin{equation}
    \underset{U\sim\mu(n,d,\mathrm{Cl}(4))}{\pr}\left[ \left|P_U[\phi] - \mathcal{U}[\phi]\right|> \tau \right] = O\left(\frac{n}{2^d\tau^2}  \right).
\end{equation}
\end{lemma}
\begin{proof}
The proof is essentially identical to the proof of Lemma~\ref{l:numerator_linear_depth}. One simply replaces the global Clifford moments from Lemma~\ref{l:Clifford_moments} with the restricted depth moments given in Lemma~\ref{l:brickwork_moments}. 
\end{proof}

For the denominator we find the same expression as in the linear depth case.
\begin{lemma}[Local random Clifford circuit output distributions are far from uniform]\label{l:pr_outside_ball}
Assume $n,d\geq 2$. Then for any $\epsilon \in [0,1/6]$,
\begin{equation}
    \underset{U\sim\mu(n,d,\mathrm{Cl}(4))}{\pr}\left[\mathrm{d}_{\mathrm{TV}}(P_U,\mathcal{U}) \geq \epsilon \right] \geq \frac{1/6-\epsilon}{1-\epsilon}.
\end{equation}
\end{lemma}
\begin{proof}
See Appendix \ref{app:pr_outside_ball}. 
\end{proof}
Once again, Eq.~\eqref{e:probability_fraction_random_circuit}, then allows us to obtain the following lower bound for the randomized statistical dimension of the decision problem of interest.

\begin{lemma}\label{l:RSD_stat_mech_bound}
For all $n$ large enough, all $d=\Omega(\log(n))$ and all $\epsilon \in [0,1/6)$ it holds that
\begin{equation}
    \mathrm{RSD}_{\tau}(\mathcal{U} \leftrightarrow\mathrm{LQC}(n,d,\mathrm{Cl}(4))\setminus\mathcal{B}(\mathcal{U},\epsilon)) = \Omega\left(\frac{\tau^22^d}{n}\right).
\end{equation}
\end{lemma}

\subsubsection{Second approach: embedding strategy}\label{sss:embedding}

\begin{figure}
\centering
\includegraphics[width=0.6\linewidth]{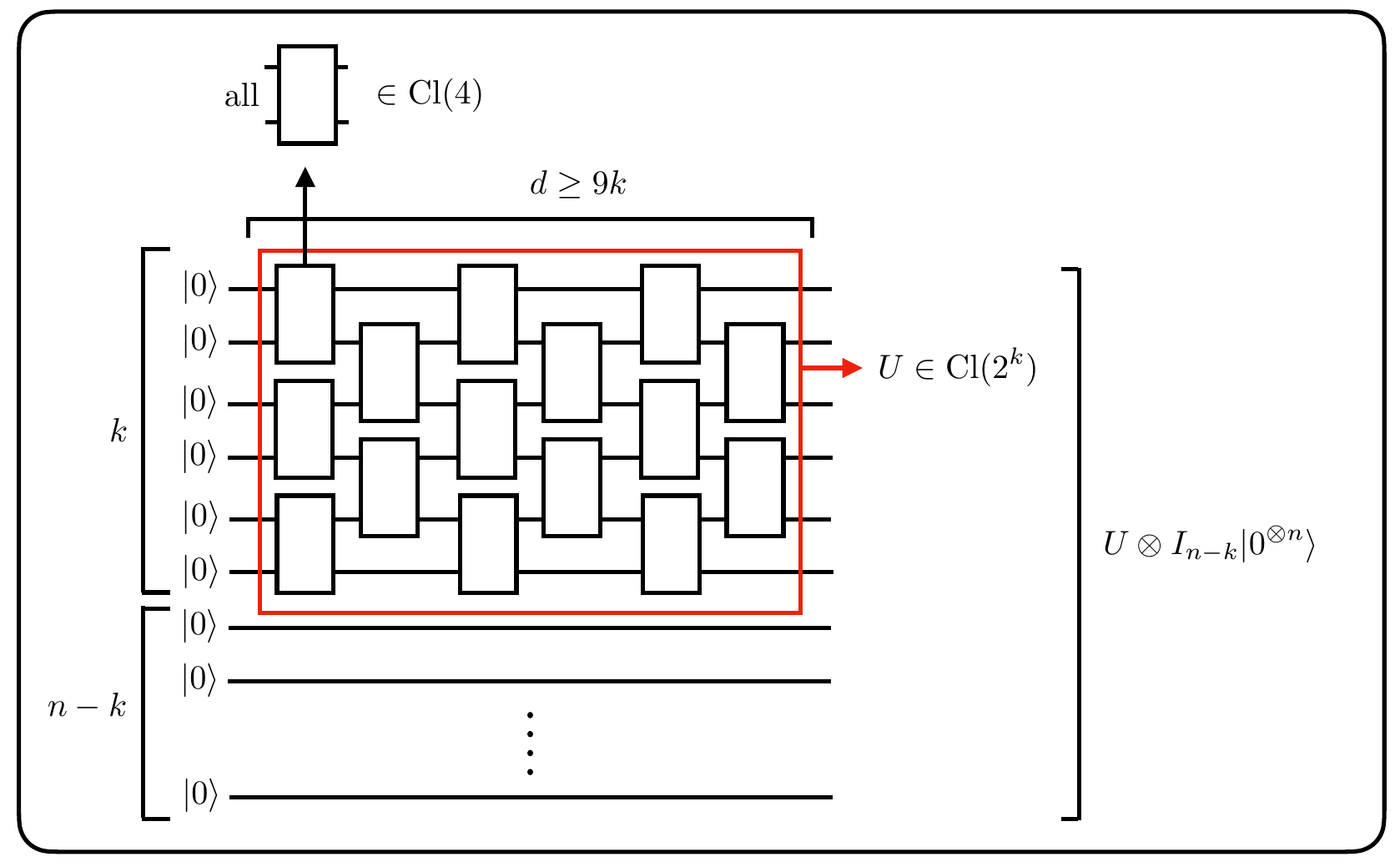}
\caption{An illustration of the local quantum circuits considered for the purpose of embedding the global-Clifford decision problem from Section~\ref{ss:linear_depth} into a subset of qubits of a sublinear depth circuit.}
\label{fig:embedding}
\end{figure}

The second approach we detail here is conceptually different. It is based on the observation that, even at sub-linear circuit depths, one can implement global Clifford unitaries on sufficiently small subsets of qubits. More specifically, for any circuit depth $d$, as long as $k \leq d/9$, one can implement any $U\in\mathrm{Cl}(2^k)$ on the first $k$ qubits of a brickwork Clifford circuit. This observation allows us to consider a decision problem with respect to $n$-qubit quantum circuits, in which we have embedded a smaller $k$-qubit version of the global-Clifford decision problem considered in Section \ref{ss:linear_depth}. Due to the nature of the embedding, the randomized statistical dimension of this decision problem will be the same as that of the global-Clifford decision problem from Section \ref{ss:linear_depth}, rescaled to $k$ qubits.

To make the argument concrete, we start by considering the subset of unitaries that arises from local nearest neighbor Clifford circuits of depth $d$, in which as illustrated in Figure \ref{fig:embedding}, only the first $k$ qubits are acted on non-trivially. More specifically, we define
\begin{equation}
	\mathrm{BQC}((k,n),d,\mathrm{Cl}(4)):=\{U\otimes I_{n-k} | U\in \mathrm{BQC}(k,d,\mathrm{Cl}(4))\}  \subseteq \mathrm{BQC}(n,d,\mathrm{Cl}(4)).
\end{equation}
Further, we denote by $\mathrm{LQC}((n,k),d,\mathrm{Cl}(4))$ the Born distributions associated with the unitaries in $\mathrm{BQC}((k,n),d,\mathrm{Cl}(4))$, i.e.
\begin{equation}
 \mathrm{LQC}((n,k),d,\mathrm{Cl}(4)) := \left\{P_U \,|\, U\in \mathrm{BQC}((k,n),d,\mathrm{Cl}(4))\right\}.
\end{equation}
Now, using the fact that $\mathrm{BQC}(k,d,\mathrm{Cl}(4)) = \mathrm{Cl}(2^k)$ whenever $d\geq 9k$, we have that
\begin{equation}
 \mathrm{LQC}((n,k),d,\mathrm{Cl}(4)) := \left\{P_{U\otimes I_{n-k}} \,|\, U\in \mathrm{Cl}(2^k)\right\},
\end{equation}
for all $k\leq d/9$. Now, let $\mathcal{U}_{(k,n)}$ be the distribution over all bit strings in $\{0,1\}^n$ whose final $n-k$ bits are all zero. Note that $\mathcal{U}_{(k,n)}$ is the Born distribution of the 1-layer circuit $H^{\otimes k} \otimes I^{\otimes (n-k)}$, where $H$ denotes the single-qubit Hadamard gate. As such, we have that $\mathcal{U}_{(k,n)}\in\mathrm{LQC}((n,k),d,\mathrm{Cl}(4)) $.
We now consider the decision problem
\begin{equation}
\mathrm{DEC}(\mathcal{U}_{(k,n)}\leftrightarrow \mathrm{LQC}((k,n),d,\mathrm{Cl}(4))\setminus \mathcal{B}(\mathcal{U}_{(k,n)},\epsilon)).\label{e:embedding_dec_problem_kn}
\end{equation}
In particular, we note that for any $k\leq d/9$ the above decision problem essentially contains the $k$-qubit global-Clifford decision problem from Section \ref{ss:linear_depth}, embedded into the first $k$ qubits. As such, we immediately obtain from Lemma \ref{l:RSD_linear_depth} that, whenever $k\leq d/9$, one has that
\begin{equation}
    \mathrm{RSD}_{\tau}(\mathcal{U}_{(k,n)} \leftrightarrow \mathrm{LQC}((k,n),d,\mathrm{Cl}(4))\setminus \mathcal{B}(\mathcal{U}_{(k,n)},\epsilon))= \Omega\left(\tau^2 2^k\right).
\end{equation}

Additionally, since the Clifford circuits restricted to the first $k$ qubits are a subset of the Clifford circuits on all $n$ qubits, we have
\begin{equation}
\mathrm{LQC}((k,n),d,\mathrm{Cl}(4))\setminus \mathcal{B}(\mathcal{U}_{(k,n)},\epsilon) \subset \mathrm{LQC}(n,d,\mathrm{Cl}(4))\setminus \mathcal{B}(\mathcal{U}_{(k,n)},\epsilon).
\end{equation}
As detailed in Observation \ref{obs:RSD_only_grows}, the RSD can only grow if the concept class grows. Hence, we conclude the validity of the following lemma.

\begin{lemma} \label{l:RSD_embedding} For all $n$ large enough, all $d>9$ and all $\epsilon \in [0,1/6]$, set $k=\left\lfloor \tfrac{d}{9} \right \rfloor$, then
\begin{equation}
\mathrm{RSD}_{\tau}(\mathcal{U}_{(k,n)} \leftrightarrow \mathrm{LQC}(n,d,\mathrm{Cl}(4))\setminus \mathcal{B}(\mathcal{U}_{(k,n)},\epsilon)) = \Omega\left(\tau^2 2^d\right).
\end{equation}
\end{lemma}

\subsubsection{Proof of Theorem \ref{t:RQCL_lower_bound}: From RSD to randomized query complexity}

In Sections \ref{sss:random_circuits} and \ref{sss:embedding} we have provided lower bounds for the randomized statistical dimension of decision problems which were designed to imply -- via Lemmas \ref{l:learning_to_decision} and \ref{l:RQC_from_RSD} -- lower bounds for learning the Born distributions of sub-linear depth brickwork Clifford circuits. We note that Theorem \ref{t:RQCL_lower_bound}, as stated in Section \ref{s:main_results}, follows directly from the results of Section \ref{sss:embedding} (as shown in the proof below). By using the results of Section \ref{sss:random_circuits} one in fact obtains a slightly weaker result, which differs from Theorem \ref{t:RQCL_lower_bound} in that (a) the randomized query complexity lower bound contains an additional factor of $1/n$ and (b) the result only holds for circuit depths $d=\Omega(\log(n))$. Neither of these differences effect the validity of Corollary \ref{c:no_sample_efficient_generic}, and as discussed before, we have provided both of these approaches in the hope of providing multiple techniques for approaching the open problems discussed in Section \ref{s:discussion}.

\begin{proof}(Theorem \ref{t:RQCL_lower_bound}) Using Lemma \ref{l:RQC_from_RSD} from Section \ref{s:prelim}, it follows from Lemma \ref{l:RSD_embedding} that for all $n$ large enough, all $d>9$ and all $\epsilon \in [0,1/6)$ 
\begin{equation}
    \mathrm{RQC}_D(\mathcal{U}_{(k,n)}, \; \mathrm{LQC}(n,d,\mathrm{Cl}(4))\setminus \mathcal{B}(\mathcal{U}_{(k,n)},\epsilon),\tau,\delta) = \Omega\left(\tau^22^d(1-2\delta)\right)\label{e:RQC_embedding}.
\end{equation}
The statement of Theorem \ref{t:RQCL_lower_bound} then follows directly by applying the reduction from deciding to learning stated in Lemma~\ref{l:learning_to_decision}.
\end{proof}

\section{Efficient learnability of Clifford circuit output distributions via samples}\label{s:clifford_learnability}

In Section \ref{s:main_results}, we have established the hardness (with respect to query complexity) of both generator-learning and evaluator-learning the output distributions of super-logarithmically deep local quantum circuits, in the statistical query model. As has been stressed in Section \ref{ss:PAC_dist}, hardness results in the SQ model \textit{do not} imply hardness results in the sample model. As such, it is of natural interest to understand whether learning the same class of distributions remains hard if one has access to a sample oracle, as opposed to a statistical query oracle. As has already been mentioned,  there are currently very few examples of computational problems which are hard in the statistical query model, but easy in the sample model~\cite{feldman2017general}. The prototypical example of such a problem is learning Boolean parity functions~\cite{kearns1998efficient}. As such, one might expect that hardness in the SQ model implies hardness in the sample model. However, we note that this intuition is most helpful when considering sufficiently \textit{unstructured} problems, where it is not clear how learning algorithms should make use of individual samples. Indeed, this is what we expect for generic local quantum circuits based on any universal gate set. However, when the problem is sufficiently well structured -- i.e., when one has a strong promise on the structure of the objects to be learned -- learning algorithms may indeed be able to exploit individual samples, and as a result such a concept class may admit an efficient learning algorithm in the sample model, even if there does not exist an efficient SQ learning algorithm. Here, we show that this is indeed the case for the output distributions of local Clifford circuits. More specifically, we show that while these distributions cannot be sample efficiently PAC learned in the SQ model, they can be both sample and computationally efficiently PAC learned in the sample model, by exploiting a strong promise on the structure of these distributions. Due to the fact that the class of distributions generated by local Clifford circuits saturates at some depth, we in fact show that the output distributions of local Clifford circuits of \textit{any depth} are computationally efficiently PAC learnable in the sample model. This class of distributions therefore provides an interesting example of a distribution concept class for which individual samples can be meaningfully exploited by learning algorithms.  

\begin{theorem}[Efficient learnability of local Clifford circuits via samples]\label{t:clifford_learn}
For all $n,d \geq 1$ the distribution concept class $\lqc(n,d,\mathrm{Cl}(4))$ is sample and computationally efficiently both PAC $\gen$-learnable and PAC $\eval$-learnable with respect to the $\sample$ oracle.
\end{theorem}
As we will see in the proof of Theorem \ref{t:clifford_learn} given in the following section, the above result relies heavily on the algebraic structure of the output states of Clifford circuits. Due to this structure, the learning problem essentially reduces to learning affine subspaces of $\mathbb{F}_2^n$. Since the set of boolean parity functions is 1-to-1 with the $n-1$-dimensional linear subspaces of $\mathbb{F}_2^n$, the learning problem at hand could hence be regarded as encoding a generalization of the prototypical problem of learning parities (although in a learning model where one can only access positive examples). Indeed, similarly to case of learning parities, the efficient learning algorithm we describe below, relies in its core on Gaussian elimination. As such, while it remains open to understand the complexity of learning the output distributions of more generic local quantum circuits in the sample model, given the hardness results of Section \ref{s:main_results} it seems natural to conjecture that the output distributions of more generic quantum circuits will remain hard to learn when moving from the SQ model to the sample model.

\subsection{Proof of Theorem \ref{t:clifford_learn}}

We prove Theorem \ref{t:clifford_learn} in three steps. Firstly, we restate known insights into the algebraic structure of stabilizer states. More specifically, we note that the Born distributions associated with these states are always the uniform distribution over some affine subspace of $\mathbb{F}^n_2$ \cite{Dehaene_2003,montanaro2017learning}. Given this we then show that, if one is given an efficient description of such an affine subspace, then it is straightforward to output either an efficient generator or evaluator for the uniform distribution over the affine subspace. As such, the problem of either generator-learning or evaluator-learning the Born distributions of stabilizer states, in the sample model, reduces to the problem of recovering an efficient description of an affine subspace of $\mathbb{F}^n_2$ when given samples from the uniform distribution over that space. We then show that this can be done efficiently by providing an efficient affine subspace recovery algorithm. We note however that our efficient affine subspace recovery algorithm is in fact a variant of a more general learning strategy called the \textit{closure algorithm}, which has previously been used to efficiently solve on-line learning problems such as learning parity functions and integer lattices 
\cite{helmbold1992learning,auer1994online}. We begin with the following lemma, synthesizing observations from 
Refs.~\cite{Dehaene_2003,montanaro2017learning}.

\begin{lemma}[Output distributions of Clifford circuits are uniform over affine subspaces \cite{Dehaene_2003,montanaro2017learning}]\label{l:output_are_u_aff}
For all $n,d\geq 1$, all elements of $\lqc(n,d,\mathrm{Cl}(4))$ are uniform distributions over an affine subspace of $\mathbb{F}^n_2$ - i.e. for all $P\in \lqc(n,d,\mathrm{Cl}(4))$ there exists an affine subspace $A_P$ of $\mathbb{F}^n_2$ such that $P= U_{A_P}$ is the uniform distribution over elements of $A_P$.
\end{lemma}

\begin{proof}
This follows directly from the fact that the output states of Clifford circuits are stabilizer states, and the observation from Refs.~\cite{Dehaene_2003,montanaro2017learning} that, up to a global phase, all $n$-qubit stabilizer state vectors $|\psi\rangle$ can be written as
\begin{equation}
    |\psi\rangle = \frac{1}{\sqrt{|A|}}\sum_{x\in A} (-i)^{l(x)}(-1)^{q(x)}|x\rangle,
\end{equation}
where $A$ is some affine subspace of $\mathbb{F}^n_2$ and $l,q$ are linear and quadratic functions on $\mathbb{F}^n_2$, respectively.
\end{proof}

Given the above result, we proceed with the following two lemmas which show that, given an efficient description of an affine subspace of $\mathbb{F}^n_2$, one can straightforwardly output both an efficient evaluator (Lemma \ref{l:eff_eval}) or efficient generator (Lemma \ref{l:eff_gen}) for the uniform distribution over the affine subspace.

\begin{lemma}[Efficient evaluation of the uniform distribution over an affine subspace]\label{l:eff_eval}
Given an affine subspace $A$ of $\mathbb{F}^n_2$ described by the tuple $(\mathbf{R},t)$, there exists an efficient evaluator for $U_A$.
\end{lemma}

\begin{proof}
The following algorithm, which simply checks whether or not a given $x\in \{0,1\}^n$ is an element of the affine subspace $A$ and then outputs the correct probability, is an evaluator for $U_A$, whose efficiency follows from the efficiency of Gaussian elimination.
\begin{algorithm}[H]
  \caption{Efficient evaluator for $U_A$
    \label{alg:efficient_evaluator}}
  \begin{algorithmic}[1]
  	\Statex Given affine subspace $A\subseteq\mathbb{F}^n_2$ via a full rank binary $n\times m$ matrix $\mathbf{R}$ and some $t\in \{0,1\}^n$, as well as some $x\in\{0,1\}^n$
    \Statex
    \State Solve $\mathbf{R}b = x-t$ via Gaussian elimination
    \If{the equation has a solution} \Comment{i.e. if $x\in A$}
        \State return $1/2^m$ 
    \Else. \Comment{i.e. if $x\notin A$}
        \State return 0
    \EndIf
  \end{algorithmic}
\end{algorithm}
\end{proof}

\begin{lemma}[Efficient generation of the uniform distribution over an affine subspace]\label{l:eff_gen}
Given an affine subspace $A$ of $\mathbb{F}^n_2$ described by the tuple $(\mathbf{R},t)$, there exists an efficient generator for $U_A$.
\end{lemma}

\begin{proof}
The following algorithm provides the desired generator.
\begin{algorithm}[H]
  \caption{Efficient generator for $U_A$
    \label{alg:efficient_generator}}
  \begin{algorithmic}[1]
  	\Statex Given affine subspace $A\subseteq\mathbb{F}^n_2$ via a full rank binary $n\times m$ matrix $\mathbf{R}$ and some $t\in \{0,1\}^n$
    \Statex
    \State Draw $b$ from the uniform distribution over $\{0,1\}^m$
    \State Output $\mathbf{R}b \oplus t$
  \end{algorithmic}
\end{algorithm}
\end{proof}

The final Lemma we require states that one can efficiently recover a description of an affine subspace of $\mathbb{F}^n_2$, when given the ability to sample from the uniform distribution over the affine subspace.

\begin{restatable}[Efficient recovery of affine subspaces]{lemma}{affinesubrecoverylemma} \label{l:affine_sub_recovery}%
Let $A$ be an affine subspace of $\mathbb{F}^n_2$. There exists an algorithm $\mathcal{A}$ which, given some $\delta\in(0,1)$, as well as access to $\sample(U_A)$, runs in time $O(\mathrm{poly}(n,1/\delta))$ and outputs, with probability at least $1-\delta$, a tuple $(\mathbf{R},t)$ which describes $A$.
\end{restatable}

\begin{proof}
See Appendix \ref{app:affine_recovery}.
\end{proof}

Finally we can now use the above Lemmas to provide a proof for Theorem \ref{t:clifford_learn}.

\begin{proof}(Theorem \ref{t:clifford_learn})
From Lemma \ref{l:output_are_u_aff} we know that for all $P\in\lqc(n,d,\mathrm{Cl}(4))$ there exists an affine subspace $A\subseteq\mathbb{F}^n_2$ such that $P=U_A$. Given this, the following algorithm provides a sample and computationally efficient PAC $\gen$-learner ($\eval$-learner) for $\lqc(n,d,\mathrm{Cl}(4))$:
\begin{enumerate}
    \item Given $\delta,\epsilon\in(0,1)$ and access to $\sample(P=U_A)$ for some unknown $P\in\lqc(n,d,\mathrm{Cl}(4))$, run the affine subspace recovery algorithm from Lemma \ref{l:affine_sub_recovery}, and receive a tuple $(\mathbf{R},t)$, which with probability at least $1-\delta$ describes $A$.
    \item Output the generator from Lemma \ref{l:eff_gen} (or evaluator from Lemma \ref{l:eff_eval}).
\end{enumerate}
For all $\delta,\epsilon \in (0,1)$, and for all $P\in\lqc(n,d,\mathrm{Cl}(4))$ the above algorithm outputs, with probability at least $1-\delta$, an \textit{exact} generator (evaluator) for $P$.
\end{proof}

\section{Discussion and Open Questions}\label{s:discussion}

 To conclude, we provide here a brief summary of our results and their implications, as well as an explicit list of open questions and directions for future research. In this work we have provided two main results. Our first result proves rigorously that the concept class of distributions obtained by measuring the output states of super-logarithmically deep local quantum circuits is not efficiently PAC learnable, by either quantum or classical learning algorithms, in the statistical query model. This result has immediate implications for the goal of proving in a rigorous way a separation between the power of quantum circuit Born machines (QCBMs) and classical probabilistic modelling techniques. More specifically, when choosing to use a QCBM based algorithm, one makes the implicit assumption that the unknown target distribution can be well approximated by a QCBM. As such, the natural set of distributions with which to try prove a separation between QCBM based algorithms and classical probabilistic modelling techniques, is indeed the class of QCBM distributions themselves -- i.e. the output distributions of local quantum circuits. To prove such a separation one requires (a) a PAC hardness result for classical learning algorithms and (b) an efficient PAC learnability result for QCBM based algorithms. However, as our first hardness result is a \textit{query complexity} lower bound, and therefore applies to both classical and quantum learning algorithms, it rules out the possibility of proving an efficient PAC learnability result for any Born machine trained via any algorithm requiring only statistical queries. As we have discussed in Appendix \ref{a:stat_query_algorithms}, many generic learning algorithms for implicit generative modelling are of this type, and as such our result provides a meaningful obstacle towards proving such separations. 

Additionally, we note that our concept class contains the output distributions of Clifford circuits, which are weakly classically simulatable - i.e. their output distributions can be efficiently sampled from given a classical description of the circuit. As such, our work also establishes that learning to sample from the output distribution of a quantum circuit given SQ oracle access to the distribution can be hard, even when such sampling can be done classically efficiently when given a circuit description. Our results therefore help to clarify the relationship between classical simulatibility of local quantum circuits, and probabilistic modelling of the Born distributions of local quantum circuits.

Finally, as there are very few known examples of computational problems which are hard in the sample model yet easy in the SQ model, hardness in the SQ model is often taken as strong evidence for hardness in the sample model. However, our second result shows that the concept class of Born distributions corresponding to local Clifford circuits is sample and time efficiently PAC learnable in the sample model by a classical learning algorithm despite being hard in the SQ model. This result therefore provides an interesting example of a probabilistic modelling problem which is hard in SQ model but efficiently PAC learnable in the sample model, and shows that, at least for highly structured distributions, one can indeed design learning algorithms which can exploit individual samples from the target distributions in a meaningful way. 

These results provide some first concrete insights into the PAC learnability of the Born distributions associated with local quantum circuits, however a variety of interesting open questions remain:\newline

\noindent \textbf{Tightness of our hardness result with respect to circuit depth:} Our main hardness result -- Theorem \ref{t:RQCL_lower_bound} -- provides a query complexity lower bound for SQ-learning the output distributions of local Clifford circuits, which scales as $\Omega{(2^d)}$, with respect to the circuit depth $d$. This implies the hardness of learning the output distributions of super-logarithmic depth Clifford circuits, but leaves open the question of whether the output distributions of logarithmic depth Clifford circuits are efficiently SQ-learnable. In order to answer this question, it is of interest to understand whether the query complexity lower bounds we have obtained in Theorem~\ref{t:RQCL_lower_bound} are \textit{tight}. In particular, can we come up with an SQ-learning algorithm which exhibits a matching upper bound $O(2^d)$ in query complexity? We note that such an algorithm would imply that the output distributions of logarithmic depth Clifford circuits \textit{are} efficiently learnable in the SQ model. We conjecture that our query complexity lower bounds are indeed tight. As we are particularly interested in understanding the efficiency of learning in the logarithmic depth regime, we formulate this conjecture as follows: 

\begin{conjecture}[SQ-learnability of log depth Clifford circuits] For all $n$, all $d=O(\log(n))$, there exists a ${\tau = \Omega(1/\mathrm{poly}(n))}$, such that the distribution concept class $\mathrm{LQC}(n,d,\mathrm{Cl}(4))$ is both sample-efficiently PAC $\gen$-learnable and $\eval$-learnable with respect to the $\mathrm{SQ}_\tau$ oracle.
\end{conjecture}

\noindent\textbf{Learnability of generic quantum circuits in the sample model:} Taken together,  our results show that the concept class of super-logarithmically deep Clifford circuits is not efficiently PAC learnable in the SQ model, but is efficiently PAC learnable in the sample model. However, the sample learnability result strongly exploits the algebraic structure of Clifford circuit output distributions, and as such it is very natural to conjecture that more generic local quantum circuits -- i.e. those with gates from $U(4)$ -- are not efficiently PAC learnable in the sample model. While our current hardness result leaves some room for proving a rigorous quantum advantage with Born machines (by considering sample based training algorithms), an analogous result in the sample model would provide a much more concrete obstacle in this regard. Once again, we make the following explicit conjecture:

\begin{conjecture}[Hardness of learning generic local quantum circuits in the sample model] For all $n$ large enough and all $d=\omega(\log(n))$, the distribution concept class $\mathrm{LQC}(n,d,\mathrm{U}(4))$ is not sample-efficiently PAC $\gen$-learnable or $\eval$-learnable with respect to the $\sample$ oracle.

\end{conjecture}

\noindent\textbf{Learnability of free-fermion distributions and match-gate circuits:} Recently, Aaronson and Grewal have investigated the learnability of passive free fermion distributions, originally claiming such distributions were efficiently learnable \cite{aaronson2021efficient}, but later retracting this claim \cite{aaronson_retract}. It is interesting to note that the learning algorithm they originally proposed uses statistical queries and as such one might be tempted to think that our techniques could straightforwardly be adapted to their setting, and used to prove the non-existence of any efficient SQ algorithm for such distributions. However, we have checked that at least for the non-number preserving version of their problem -- which corresponds to the output distributions of local quantum circuits with gates from the matchgate group -- this straightforward adaptation fails to provide super-polynomial lower bounds. Thus the question of whether or not free fermion distributions are learnable remains open. As in this work, it may be interesting as a first step to investigate learnability within the statistical query model, via lower bounds on the randomized statistical dimension.\newline

\noindent\textbf{Average-case hardness of PAC learning:} The SQ PAC hardness result we have obtained is of a worst-case nature -- i.e. it implies that for all efficient learning algorithms there exists at least one distribution in the concept class for which the learning algorithm cannot succeed. However, it would also be interesting to understand the \textit{average-case} PAC learnability of local quantum circuit output distributions. More specifically, what can one say if we relax the requirement that \textit{all} distributions in the concept class be efficiently learnable to a requirement that with some fixed probability, a randomly drawn distribution from the concept class will be efficiently PAC learnable? From a practical perspective this is perhaps the more interesting question, as worst-case instances may not correspond to the distributions encoding practically-relevant problems. Additionally, average-case hardness results would also allow one to better understand the limitations of \textit{heuristic} probabilistic modelling algorithms. \newline

\noindent \textbf{Robustness of learnability result with respect to noise:} The sample-learnability result we have provided in Section \ref{s:clifford_learnability} for the output distributions of Clifford circuits relies on (a) a promise that the output distribution is the uniform distribution over some affine subspace, and (b) the existence of an efficient affine subspace recovery algorithm. As the promise (a) does not hold in the presence of realistic noise models, our learnability result does not immediately hold for the output distributions of noisy local Clifford circuits. As in practice any such quantum circuit will be noisy, it is of great practical interest to understand whether the efficient learning algorithm we provide can be made robust to realistic noise, possibly through the use of robust subspace recovery techniques. Alternatively, the SQ model was originally introduced as a way of obtaining noise-tolerant learning algorithms for Boolean functions. However, in the probabilistic modelling setting it is not yet clear to which extent efficient SQ algorithms imply noise-robust learning algorithms. In light of this, it would be interesting to formalize this relationship, in which case any efficient SQ algorithm for Clifford circuits of depth $d=O(\log(n))$ might immediately give rise to a noise tolerant algorithm, for some specific class of noise models. Additionally, if one can provide a noise-robust sample-based algorithm for super-logarithmically deep Clifford circuits, this would provide an interesting example of a distribution class admitting an efficient noise-robust learning algorithm, but not an efficient SQ learning algorithm. \newline 

\noindent\textbf{Implications of our result in other areas:} In other settings it is known that hardness results for the SQ learnability of certain concept classes, and the associated lower bounds on the randomized statistical dimension of certain decision problems, have meaningful implications for open questions in areas such as communication complexity, property testing and learning with privacy or robustness guarantees \cite{feldman2017general,diakonikolas2017statistical}. In light of this, it would be of interest to understand whether our key technical result, a lower bound on the randomized statistical dimension of a local quantum circuit based decision problem, can be leveraged to obtain insights into open problems in quantum communication complexity, quantum verification or private quantum learning.

\appendix

\begin{acknowledgements}

\noindent We gratefully thank Paul Boes, Nana Liu and the QML reading group at the FU Berlin for many insightful discussions. 
The Berlin team thanks the BMWi (PlanQK), the DFG (project B01 of CRC 183 and EI 519/21-1), the Einstein Foundation,
the BMBF (Hybrid) and the MATH+ cluster  of excellence for support. Y.~Q.~acknowledges the support of an NUS Overseas Graduate Fellowship and a Stanford QFARM PhD fellowship.
D.H.\ acknowledges funding from the U.S.\ Department of Defense through a QuICS Hartree fellowship. 
\end{acknowledgements}

\section*{Author Contributions}
\noindent The following describes the different contributions of all authors of this work, using roles defined by the CRediT (Contributor Roles Taxonomy) project \cite{credit}: \newline

\noindent \textbf{MH:} Conceptualization, Formal Analysis, Methodology, Writing (Original Draft), \textbf{MI:} Conceptualization, Formal Analysis, Methodology, Writing (review and editing), \textbf{AN:} Conceptualization, Formal Analysis, Methodology, Writing (review and editing), \textbf{JH:} Formal Analysis, Writing (review and editing), \textbf{YQ:} Conceptualization, Writing (review and editing), \textbf{DH:} Conceptualization, Writing (review and editing), \textbf{JPS:} Conceptualization, Writing (review and editing), \textbf{JE:} Conceptualization, Writing (review and editing), \textbf{RS:} Conceptualization, Formal Analysis, Methodology, Supervision, Writing (Original Draft)

\section{On the applicability of SQ hardness results for practical generative modelling algorithms}\label{a:stat_query_algorithms}

Given that (a) one of our main motivations in studying the learnability of local quantum circuit Born distributions is to understand the potential for obtaining a meaningful quantum advantage via QCBM based generative modelling algorithms and (b) our primary hardness result applies only within the statistical query model, it is of interest to understand the extent to which our results apply to practically utilized generative modelling algorithms. Conveniently, Mohamed and Lakshminarayanan have recently provided an excellent review of generic learning algorithms for implicit generative models \cite{mohamed2017learning}, and our aim in this section is to provide a brief insight into the observation that indeed almost all of the algorithms they consider -- which include algorithms used for training QCBMs \cite{coyle2020born} -- can be efficiently executed in the SQ model, and are therefore under the domain of applicability of our hardness result. We stress that our goal here is \textit{not} to provide a detailed review of algorithms for generative modelling from the statistical query perspective, but rather to provide the tools necessary to convince oneself that many of the algorithms reviewed in Ref. \cite{mohamed2017learning} can indeed be efficiently executed with access only to a statistical query oracle for the unknown target distribution. More specifically, we first note that most of the algorithms from Ref. \cite{mohamed2017learning} rely only on the expectation values of function outputs. We then  focus on showing how algorithms which require the expectation values of functions with arbitrary but constant codomains, or multiple inputs, can be efficiently simulated with access to a standard statistical query oracle, as defined in Section \ref{s:prelim}. Additionally, it is important to stress that one cannot hope to claim that \textit{all} generative modelling algorithms can be efficiently simulated in the SQ model. Indeed, in Section \ref{s:clifford_learnability} we have given a generative modelling algorithm which is able to use a promise on the structure of the concept class to exploit individual samples from the target distribution. Rather, our focus here is on the \textit{generic} generative modelling algorithms, which are often used in practice for the optimization of state-of-the-art generative models such as GANs and QCBMs.

Typically, when designing learning algorithms for implicit generative models, one assumes $\sample$ access to some unknown target distribution $Q$, as well as access to a parameterized generator $\gen(\btheta)$ which can be used to generate samples from the corresponding model distribution $P_{\btheta}$. For example, when the generator is a QCBM one has that
\begin{equation}
    P_{\btheta}(x) = |\langle x|U(\btheta)|0\rangle^2,
\end{equation}
and samples from $P_{\btheta}$ are obtained by measuring the state vector $U(\btheta)|0\rangle$ in the computational basis. When the generator is some classical parameterized function $f_{\btheta}:\{0,1\}^m \rightarrow \{0,1\}^n$ (which could be a neural network) one has that
\begin{equation}
    P_{\btheta}(x) = \frac{1}{2^m}\sum_{y\in\{0,1\}^m} \delta(f_{\btheta}(y),x),
\end{equation}
where $\delta(x,y) = 1$ if $x=y$ and $\delta(x,y) = 0$ otherwise, and one generates samples from $P_{\btheta}$ by first drawing $y\in\{0,1\}^m$ uniformly at random and then outputting $f(y)$.
The goal is then to identify a suitable set of parameters $\boldsymbol{\theta}^*$ such that the model distribution $P_{\btheta}$ is sufficiently close to the target distribution $Q$. In order to do this, one needs a method via which to compare the target distribution $Q$ with the current model distribution $P_{\btheta}$. As discussed in detail in Ref. \cite{mohamed2017learning}, in order to do this, one typically begins by constructing a parameterized estimator $r_{\bphi}$ of either the \textit{density difference} $Q-P_{\btheta}$ or the \textit{density ratio} $Q/P_{\btheta}$. Given this, one then defines a loss function $\mathcal{L}(\btheta,\bphi)$, after which learning proceeds by alternating optimization -- typically via gradient descent type algorithms -- of the loss with respect to comparison parameters $\bphi$ and model parameters $\btheta$. Given this generic framework, the question we are interested in here is to what extent one can evaluate the loss functions (and their gradients) of practical learning algorithms, using \textit{only statistical query access to the target distribution}~$Q$.

Perhaps surprisingly, we find that almost all the loss functions discussed in Ref. \cite{mohamed2017learning} are defined via the expectation values of (possibly parameterized) functions with respect to both the model and target distributions, and do not rely on direct comparisons of individual samples. In practice these loss functions are then estimated via sample mean estimates or Monte Carlo methods, and as such it seems promising that one could evaluate these loss functions with access to only an SQ oracle for the target distribution. In order to see that this is indeed the case, we have to take care of two subtleties. The first issue is that we have defined the SQ oracle as an oracle which can be queried on functions $g:\mathcal{X}\rightarrow [-1,1]$. However, many of the functions whose expectations are required for the loss functions discussed in Ref. \cite{mohamed2017learning} have a different codomain, or may even be unbounded (as in the case of functions which estimate the density ratio for example). The second issue is that many of the functions discussed in Ref.~\cite{mohamed2017learning} are of the type $g:\mathcal{X}\times\mathcal{X}\rightarrow [-1,1]$, and one requires the expectation value of the function output with respect to inputs drawn both from the model and the target distribution. 

We begin with a discussion of the first issue - that of a a single input, but an alternative codomain. While it is straightforward to define a generalization of the SQ oracle which can be queried via functions with a different codomain, it \textit{is not} immediately apparent that hardness results with respect to the original SQ oracle, also apply to learning algorithms given access to such a generalized SQ oracle. Luckily however, as we show below, this is indeed the case for any SQ oracle with a \textit{constant} finite-interval codomain. As in practice any potentially unbounded loss function would be truncated to some fixed interval, it is sufficient to restrict ourselves to such intervals. We begin by defining a generalized SQ oracle.

\begin{definition}[Generalized SQ oracle] Given $P\in\mathcal{D}_n$, some $\tau\in[0,1]$ and some non-zero interval $[a,b]$, define $\SQ^{[a,b]}_\tau[P]$ as the oracle which, when queried via some function $\phi:\{0,1\}^n\rightarrow[a,b]$ responds with some $v$ such that $|P[\phi]-v|\leq \tau$.
\end{definition}
We note that in this notation, the standard SQ oracle -- as defined in Definition \ref{d:oracles} -- is an $\SQ^{[-1,1]}_{\tau}$ oracle. Given this definition, we then have the following theorem, which shows that the randomized query complexity of a generative modelling problem with respect to generalized SQ oracle, can be lower bounded by the randomized query complexity with respect to the standard SQ oracle, but with the tolerance rescaled \textit{by a constant factor}.
\begin{theorem}[Generalized SQ query complexity lower bounds via tolerance rescaling]\label{t:range_invariance_SQ}
For all distribution concept classes $\mathcal{C}\subseteq\mathcal{D}_n$, all $\epsilon,\delta,\tau\in(0,1)$, and all constants $a<b$, one has that
\begin{equation}
    \mathrm{RQC}_L(\mathcal{C},\SQ^{[a,b]}_\tau,\delta,\epsilon,\gen) \geq \mathrm{RQC}_L(\mathcal{C},\SQ^{[-1,1]}_{2\tau/(b-a)},\delta,\epsilon,\gen).
\end{equation}
\end{theorem}
\begin{proof}  Assume there exists some algorithm $\mathcal{A}_1$ which, when given access to $\SQ^{[a,b]}_\tau[P]$, uses $M$ queries ${\{\phi_i\,|\, i \in [M]\}}$ to $\SQ^{[a,b]}_\tau[P]$ before outputting a generator. We will construct an algorithm $\mathcal{A}_2$, which when given access to $\SQ^{[-1,1]}_{2\tau/(b-a)}[P]$, uses the same number of queries and outputs the same generator. To do this, we start by defining the function $f:[-1,1]\rightarrow [a,b]$ via
\begin{equation}
    f(x) = x\left(\frac{b-a}{2}\right) + \frac{a+b}{2}.
\end{equation}
We note that $f$ is invertible, and that
\begin{equation}
    x\in[u-v,u+v]\implies f(x)\in \left[f(u) - v\left(\frac{b-a}{2}\right), f(u) + v\left(\frac{b-a}{2}\right)\right].
\end{equation}
Now, when given access to $\SQ^{[-1,1]}_{2\tau/(b-a)}[P]$, algorithm $\mathcal{A}_2$ does the following:
\begin{enumerate}
    \item Runs $\mathcal{A}_1$ and receives its first query function $\phi_1:\mathcal{X}\rightarrow[a,b]$.
    \item Queries $\SQ^{[-1,1]}_{2\tau/(b-a)}[P]$ on $\hat{\phi}_1 = f^{-1}\circ \phi_1$ and receives some $\hat{y}_1\in [-1,1]$.
    \item Sends $y_1 = f(\hat{y}_1)$ to $\mathcal{A}_1$.
    \item If $\mathcal{A}_1$ outputs a generator then $\mathcal{A}_2$ outputs the same generator. If $\mathcal{A}_1$ wants to make another query $\phi_i$, then repeat steps (1) to (3).
\end{enumerate}
By virtue of the properties of $f$ and $\SQ^{[-1,1]}_{2\tau/(b-a)}[P]$ we note that for all $i$, we have that 
\begin{equation}
    y_i \in [P[\phi_i]-\tau,P[\phi_i] + \tau]
\end{equation}
and as a result algorithm $\mathcal{A}_1$ cannot distinguish what it receives from algorithm $\mathcal{A}_2$ from what it would have received by directly querying $\SQ^{[a,b]}_{\tau}[P]$. As such, algorithm $\mathcal{A}_2$, when given access to to $\SQ^{[-1,1]}_{2\tau/(b-a)}[P]$, will output the same generator, after the same number of queries, as algorithm $\mathcal{A}_1$ would when given access to $\SQ^{[a,b]}_{\tau}[P]$.
\end{proof}
As a corollary of the above theorem, we see that the asymptotic randomized query complexity of any learning algorithm can only increase when going from a standard SQ oracle to a generalized SQ oracle, as a result of the fact that the constant factor rescaling of the tolerance $\tau$ is irrelevant from the asymptotic perspective. In particular our main hardness result -- Theorem \ref{t:RQCL_lower_bound} -- holds even for generalized SQ oracles, and therefore applies to algorithms which require expectation values of functions with arbitrary constant codomains.

\begin{corollary}[From Theorem \ref{t:RQCL_lower_bound} and Theorem \ref{t:range_invariance_SQ}] For all $n,d$ large enough, all constants $a<b$ and for all $\epsilon \in [0,1/6]$
\begin{equation}
    \mathrm{RQC}_L(\mathrm{LQC}(n,d,\mathrm{Cl}(4)),\mathrm{SQ}^{[a,b]}_{\tau},\delta,\epsilon,\gen) =\Omega\left(\tau^22^d(1-2\delta)
    \right).
\end{equation}
\end{corollary}
The second issue we need to address is the fact that many of the loss functions discussed Ref. \cite{mohamed2017learning} contain terms of the following type
\begin{equation}
    \mathbb{E}_{x\sim P_{\btheta}, y\sim Q}[K(x,y)],
\end{equation}
for functions $K:\mathcal{X}\times\mathcal{X}\rightarrow[a,b]$. This is for example the case when using the \textit{maximum mean discrepancy} as a loss function~\cite{mmd}. For example, when using the maximum mean discrepancy for training quantum circuit Born machines, the kernel $K$ is typically taken to be the symmetric Gaussian kernel
\begin{equation}
    K(x,y) = \frac{1}{c}\sum_{j = 1}^c\mathrm{exp}\left(-\frac{1}{2\sigma_i}||x-y||_2^2\right),
\end{equation}
where $\{\sigma_i\}$ is a set of ``bandwidths" \cite{coyle2020born,Liu_2018}. Of course one cannot evaluate this expectation value  exactly, and so typically one uses $\sample(P_{\btheta})$ and $\sample{(Q)}$ access to construct a suitable \textit{estimator} . More specifically, typically one draws $m$ samples $\{x_i\}$ from $P_{\btheta}$, and $n$ samples $\{y_j\}$ from $Q$, which allows one to use the unbiased sample-mean estimator
\begin{equation}
    \hat{E}_{m,n} = \frac{1}{mn}\sum_{i = 1}^n\sum_{j = 1}^nK(x_i,y_j).
\end{equation}
However, a brief moment of thought shows that one can construct the same estimator with access to a sufficiently accurate $\SQ$ oracle for $Q$. In particular, define the function $K_x := K(x,\cdot)$, and note that 
\begin{equation}
    \hat{E}_{m,n} = \frac{1}{m}\sum_{i = 1}^n\left(\frac{1}{n}\sum_{j = 1}^nK_{x_i}(y_j)\right)\\
    =\frac{1}{m}\sum_{i = 1}^n\hat{E}_{m}(K_{x_i})
\end{equation}
where $\hat{E}_{m}(K_{x_i})$ is the sample mean estimator of $\mathbb{E}_{y\sim Q}[K_{x_i}(y)]$. One can therefore simply replace  sample-mean estimate of $\mathbb{E}_{y\sim Q}[K_{x_i}(y)]$ with an SQ query to $\SQ_\tau[Q]$ -- i.e. one can use the estimator
\begin{equation}
    \hat{E}_{m,\mathrm{SQ}} = \frac{1}{m}\sum_{i = 1}^m \query[\mathrm{SQ}_\tau](K_{x_i}).
\end{equation}
In summary, we see that while the SQ oracle was originally defined with respect to functions $f:\mathcal{X}\rightarrow[-1,1]$, our hardness results apply also to algorithms which require expectation values of functions with alternative but constant codomains, or multiple inputs. As such, our hardness results are indeed applicable to many practically used state of the art generative modelling algorithms surveyed in Ref. \cite{mohamed2017learning}.

\section{Proof of Lemma \ref{l:learning_to_decision}}\label{app:l_to_d}

\learningimpliesdeciding*

\begin{proof} Given some $n,\epsilon,\tau,\delta,\mathcal{C}$ and $D_0$ satisfying the assumptions of the lemma, let us denote by $\mathcal{A}_L$ the  randomized $\SQ_\tau$-PAC $\gen$-learner ($\eval$-learner) for $\mathcal{C}$ which achieves query complexity $\mathrm{RQC}_L(\mathcal{C},\mathrm{SQ}_\tau,\delta,\epsilon,\gen (\eval))$. We now define an algorithm $\mathcal{A}_D$, which solves the decision problem $\mathrm{DEC}(D_0 \leftrightarrow\mathcal{C}\setminus\mathcal{B}(D_0,\epsilon))$ with $\mathrm{SQ}_\tau$, using $\mathrm{RQC}_L(\mathcal{C},\mathrm{SQ}_\tau,\delta,\epsilon)$ queries (for convenience, from this point on we drop the $\gen$ and $\eval$ indicators from the notation for randomized query complexity, as it is clear from the context which learner we are referring to).  In particular, on input $\delta$, and given access to $\SQ_{\tau}(P)$, algorithm $\mathcal{A}_D$ does the following:
\begin{enumerate}
    \item Use $\mathcal{A}_L$, with inputs $\delta$ and $\epsilon/2$, and obtain, with probability at least $1-\delta$, a generator $\gen_Q$ (or evaluator $\eval_Q$) for some $Q$ satisfying $\mathrm{d}_{\mathrm{TV}}(P,Q) \leq \epsilon/2$.
    \item Use $\gen_Q$ (or $\eval_Q$) to calculate $\mathrm{d}_{\mathrm{TV}}(Q,D_0)$. This can be done, potentially inefficiently, without access to additional samples.
    \begin{enumerate}
        \item If $\mathrm{d}_{\mathrm{TV}}(Q,D_0) \leq \epsilon/2$ output 1.
        \item If $\mathrm{d}_{\mathrm{TV}}(Q,D_0) > \epsilon/2$ output 0.
    \end{enumerate}
\end{enumerate}
By construction, we have that if $P=D_0$, then with probability at least $1-\delta$, algorithm $\mathcal{A}_D$ will output 1, and if ${P\in \mathcal{C}\setminus\mathcal{B}(U,\epsilon)}$, then with probability at least $1-\delta$ algorithm $\mathcal{A}_D$ will output 0. As a result, $\mathcal{A}_D$ indeed solves the decision problem ${\mathrm{DEC}(D_0 \leftrightarrow\mathcal{C}\setminus\mathcal{B}(U,\epsilon))}$ with $\SQ_\tau$. As $\mathcal{A}_L$ is assumed to have query complexity $\mathrm{RQC}_L(\mathcal{C},\mathrm{SQ}_\tau,\delta,\epsilon)$, algorithm $\mathcal{A}_D$ will also have query complexity $\mathrm{RQC}_L(\mathcal{C},\mathrm{SQ}_\tau,\delta,\epsilon)$, and therefore $\mathrm{RQC}_L(\mathcal{C},\mathrm{SQ}_\tau,\delta,\epsilon)$ provides an upper bound on ${\mathrm{RQC}_D(D_0,\mathcal{C}\setminus\mathcal{B}(D_0,\epsilon),\tau,\delta)}$.
\end{proof}

\section{Moment calculations}\label{app:moments}
In this appendix we provide proofs for Lemmas \ref{l:Clifford_moments} and \ref{l:brickwork_moments} from the main text, which provided expressions for moments with respect to random global Clifford unitaries and random sublinear depth Clifford circuits respectively. In order to prove these Lemma's we require some preliminaries. We start by introducing the concept of a unitary $t$-design:
\begin{definition}[Unitary $t$-design]\label{d:design}
Let $\mu$ be an ensemble of unitary operators  $U\in\mathrm{U}(D)$ on $\mathbb{C}^D$. Then, $\mu$ is a unitary $t$-design if, for every polynomial $f(U,\bar{U})$ of degree at most $t$ in the matrix elements of $U$ and degree at most $t$ in the matrix elements of the complex conjugate $\bar{U}$, it holds that
\begin{equation}
\underset{U\sim \mu}{\mathbb{E}} \left[ f(U,\bar{U})\right] =\underset{U\sim \mathrm{U}(D)}{\mathbb{E}} \left[ f(U,\bar{U}) \right]
\end{equation}
where $U\sim \mathrm{U}(D)$ denotes $U$ being drawn at random from the Haar measure over the unitary group $\mathrm{U}(D)$.
\end{definition}
A very prominent example of a unitary design is given by the Clifford group:
\begin{theorem}[\cite{webb2016Clifford, zhu2017multiqubit}]
The uniform measure over the Clifford group $\mathrm{Cl}(2^n)$ is an exact unitary 3-design for all $n$.
\label{thm:Clifford_group_3_design}
\end{theorem}
As a consequence, we can replace all first, second, and third moments with respect to the Clifford group with the corresponding moments with respect to the Haar measure. Unitary $t$-designs are very useful since expectation values with respect to the Haar measure can be evaluated analytically. For our purposes, the following special formula is sufficient:

\begin{lemma}[Moments from $t$-designs, Lemma 2.2.2 in Ref. \cite{low2010pseudorandomness}]\label{l:t_moments}
Let $\mu$ be an ensemble of unitary operators  $U\in\mathrm{U}(2^n)$ on $\left(\mathbb{C}^2 \right)^{\otimes n}$. Let $\mu$ form a unitary $t$-design, then 
\begin{equation}
    \underset{U \sim \mu}{\mathbb{E}}\left[ (U \ket{0^{\otimes n}}\bra{0^{\otimes n}} U^{\dagger})^{\otimes t}\right ] = \frac{\sum_{\pi \in S_t}W_{\pi}}{t!\binom{2^n+t-1}{t}}
\end{equation}
where $S_t$ is the symmetric group on $t$ elements, and $W_{\pi}$ is the permutation operator on the $t$-fold tensor product corresponding to the permutation. $\pi\in S_t$ 
\end{lemma}

Given these preliminaries we can now prove Lemmas \ref{l:Clifford_moments} and \ref{l:brickwork_moments}, which we restate here for convenience.

\cliffordmoments*
\begin{proof}
Note that we are treating a first and a second moment with respect to the Clifford group $\mathrm{Cl}(2^n)$. By the 3-design property of the Clifford group (see Theorem \ref{thm:Clifford_group_3_design}), these moments coincide with those with respect to the Haar measure. For the first moment, we therefore have
\begin{align}
  \underset{U\sim \mathrm{Cl}(2^n)}{\mathbb{E}} \left[ P_U(x)  \right]
    &= \underset{U\sim \mathrm{U}(2^n)}{\mathbb{E}} \mathrm{Tr}\left(\ket{x}\bra{x} U \ket{0^{\otimes n}}\bra{0^{\otimes n}}U^{\dagger}\right)\\
    &= \mathrm{Tr}\left(\ket{x}\bra{x} \underset{U\sim \mathrm{Cl}(2^n)}{\mathbb{E}}\left[U \ket{0^{\otimes n}}\bra{0^{\otimes n}}U^{\dagger}\right]\right)\\
    &= \frac{1}{2^n}\mathrm{Tr}\left(\ket{x}\bra{x} \mathds{1}\right) = \frac{1}{2^n},
\end{align}
where, when going from the second to the third line, the Haar expectation value was evaluated via Lemma \ref{l:t_moments}. For the second moment, we have
\begin{align}
  \underset{U\sim \mathrm{Cl}(2^n)}{\mathbb{E}} \left[ P_U(x) P_U(y) \right]
    &= \underset{U\sim \mathrm{U}(2^n)}{\mathbb{E}} \mathrm{Tr}\left(\ket{x}\bra{x} U \ket{0^{\otimes n}}\bra{0^{\otimes n}}U^{\dagger}\right)\mathrm{Tr}(\ket{y}\bra{y} U \ket{0^{\otimes n}}\bra{0^{\otimes n}}U^{\dagger})\\
    &= \mathrm{Tr}\left((\ket{x}\bra{x} \otimes \ket{y}\bra{y}) \underset{U\sim \mathrm{U}(2^n)}{\mathbb{E}} (U^{\otimes 2} \left(\ket{0^{\otimes n}}\bra{0^{\otimes n}}\right)^{\otimes 2}U^{\dagger \otimes 2})\right)\\
    &=\frac{1}{2^n(2^n+1)}\mathrm{Tr}\left((\ket{x}\bra{x} \otimes \ket{y}\bra{y})(\mathbb{I}+\mathbb{F})\right)\\
    &=\frac{1}{2^n(2^n+1)}[1 + |\langle x| y\rangle|^2]\\
    &=\frac{1}{2^n(2^n+1)}[1 + \delta_{x,y}],
\end{align}
where $\mathbb{F}$ denotes the flip operator acting as $\mathbb{F}(\ket{x}\otimes\ket{y})=\ket{y}\otimes\ket{x}$.
\end{proof}

\restricteddepthcliffordmoments*

\begin{proof}
We note that $\mu(n,d,\mathrm{Cl}(4))$ is an exact 1-design at any depth $d$\footnote{In fact, already a single layer of randomly drawn Clifford gates, i.e. $\mu(n,d=1,\mathrm{Cl}(4))$, forms an exact 1-design. This is because this layer contains as a subgroup the Pauli group which is known to form an exact 1-design. It follows from the invariance of the Haar measure under left multiplication that random Clifford circuits form an exact 1-design also for $d\geq1$.}. Hence, the first moment is the same as in Eq. \eqref{eq:Clifford_first_mom} for the full Clifford group, i.e.
\begin{equation}
 \underset{U\sim \mu(n,d,\mathrm{Cl}(4))}{\mathbb{E}} \left[ P_U(x) \right] 
	= \underset{U\sim \mathrm{U}(2^n)}{\mathbb{E}} \left[ P_U(x) \right] = \frac{1}{2^n}.
\end{equation}
To obtain the second moment given in Eq. \eqref{eq:brickwork_second_mom}, we adapt and modify a calculation presented in Section 6.3 of Ref.~\cite{barak2021spoofing}.  Specifically, using a mapping to a statistical mechanics model, the second moment with respect to the random circuit, $\underset{U\sim \mu(n,d,\mathrm{Cl}(4))}{\mathbb{E}} \left[ P_U(x) P_U(y) \right] $, can be expressed as a partition function. The value of this partition function can then be bounded by counting domain walls. In Section 6.3 of Ref. \cite{barak2021spoofing}, this technique was already used to obtain an upper bound on $\underset{U\sim\mu(n,d,\mathrm{Cl}(4)}{\mathbb{E}} \left[ P_U(x)^2 \right]$, for random circuits of depth $d=\Omega(\log n)$. More specifically, Ref. \cite{barak2021spoofing} has obtained the upper bound

\begin{equation}
	\underset{U\sim\mu(n,d,\mathrm{Cl}(4)}{\mathbb{E}} \left[ P_U(x)^2 \right]\leq \left(1+\left(\frac{4}{5}\right)^{d}\right)^{n/2} \underset{U\sim \mathrm{U}(2^n)}{\mathbb{E}} \left[ P_U(x)^2 \right],
\end{equation}
which is given in terms of the Haar expectation value $\underset{U\sim \mathrm{U}(2^n)}{\mathbb{E}} \left[ P_U(x)^2 \right]$, and indeed converges to this Haar value in the infinite circuit depth-limit $d\to \infty$. 
A similar analysis allows us to obtain the following bound on the expectation value of the cross terms $P_U(x) P_U(y)$,
\begin{equation}
	\underset{U\sim\mu(n,d,\mathrm{Cl}(4))}{\mathbb{E}} \left[ P_U(x) P_U(y) \right] \leq \left(1+\left(\frac{4}{5}\right)^{d}\right)^{n/2} \underset{U\sim \mathrm{U}(2^n)}{\mathbb{E}} \left[ P_U(x) P_U(y) \right].\label{e:clif_via_haar}
\end{equation}
Note that this upper bound is also given in terms of the corresponding Haar value $\underset{U\sim \mathrm{U}(2^n)}{\mathbb{E}} \left[ P_U(x) P_U(y)\right]$. As the second moments with respect to the uniform measure over the Clifford group coincide with those of the Haar measure over the full unitary group, we can use the second moment already calculated in Lemma \ref{l:Clifford_moments}. Finally, we bound the prefactor: 
By Bernoulli's inequality, we have that $(1+x^d)^n \leq e^{nx^d}$. For $d=\Omega(\log n)$ and $x<1$ we can then use the convexity of the exponential function $e^{y}\leq(1-y)e^0+ye^1$ to obtain $e^{nx^d}\leq 1-nx^d + enx^d\leq 1+2nx^d$. This allows us to show that 
\begin{equation}
	\left(1+\left(\frac{4}{5}\right)^{d}\right)^{n/2} \leq 1 + n \left(\frac{4}{5}\right)^{d}.\label{eq:prefactor}
\end{equation}

\noindent Substituting Eqs. \eqref{eq:Clifford_second_mom} and \eqref{eq:prefactor} into Eq. \eqref{e:clif_via_haar} then yields Eq. \eqref{eq:brickwork_second_mom}.

\end{proof}

\section{Proof of Lemma \ref{l:pr_outside_ball_linear_depth} and \ref{l:pr_outside_ball}} \label{app:pr_outside_ball}
In this section, we will prove Lemma \ref{l:pr_outside_ball_linear_depth} and Lemma \ref{l:pr_outside_ball} which we jointly restate as follows:

\begin{lemma}[Restatement of Lemmas \ref{l:pr_outside_ball_linear_depth} and \ref{l:pr_outside_ball}]\label{l:joint_prob_outside_ball}
Assume $n,d\geq 2$. Let $\mu=\mathcal{U}_{\mathrm{Cl}(2^n)}$ or $\mu=\mu(n,d,\mathrm{Cl}(4))$. Then, for any $\epsilon \in [0,1/6]$, it holds that
\begin{equation}\label{e:clif_measurement_differences}
    \underset{U\sim \mu}{\pr} \left[\mathrm{d}_{\mathrm{TV}}(P_U,\mathcal{U}) \geq \epsilon \right] \geq \frac{1/6-\epsilon}{1-\epsilon}.
\end{equation}
\end{lemma}
In order to prove this lemma, we start with the following observation.
\begin{lemma}\label{l:clifford_single_qubit}
Let $\mathrm{Cl}(2)\subset \mathrm{U}(2)$ denote the single-qubit Clifford group. Then
\begin{equation}
   \underset{U\sim\mathrm{Cl}(2)}{\mathbb{E}}[|P_U(0)-P_U(1)|] = \frac{1}{3}. 
\end{equation}
\end{lemma}
\begin{proof}
Let $\mathrm{Stab}(n)$ be the set of $n$-qubit stabilizer states, and note that for any function $f$ on $n$-qubit states
\begin{equation}
    \underset{U\sim\mathrm{Cl}(2^n)}{\mathbb{E}}\left[f\left(C|0\rangle\right)\right]=
    \underset{\psi\sim\mathrm{Stab}(n)}{\mathbb{E}}_{}\left[f\left(|\psi\rangle\right)\right].
\end{equation}
This is because drawing random Cliffords and applying them to $\ket 0$ induces a distribution on states which is precisely the uniform distribution over stabilizer states. In particular, we then have that
\begin{align}
    \underset{U\sim\mathrm{Cl}(2)}{\mathbb{E}}[|P_U(0)-P_U(1)|]&=
    \underset{\psi\sim\mathrm{Stab}(1)}{\mathbb{E}}\left[\left|\left|\langle 0| \psi\rangle \right|^{2}-\left|\langle 1| \psi\rangle\right|^{2}\right|\right] \\
    &:= \underset{\psi\sim\mathrm{Stab}(1)}{\mathbb{E}} \left|P_{\ket{\psi}}\left(0\right)-P_{\ket{\psi}}\left(1\right)\right|.
\end{align}
There are six stabilizer state vectors on 
a single qubit, namely $ \mathrm{Stab}(1) = \left\{ \ket 0,\ket 1,\ket +,\ket -,\ket i,\ket{-i}\right\}$ (the eigenvectors
of the Pauli matrices $Z,X,Y$, respectively). Note that $\ket +,\ket -,\ket i,\ket{-i}$ all lead to the uniform distribution with respect to computational basis measurements, whereas $\ket 0$,$\ket 1$ lead to totally biased output distributions. Averaging uniformly over those 6 states, we find
\begin{align}
    \underset{\psi\sim\mathrm{Stab}(1)}{\mathbb{E}} \left|P_{\ket{\psi}}\left(0\right)-P_{\ket{\psi}}\left(1\right)\right|
    &=\sum_{\ket{\psi}\in \mathrm{Stab}(1)}\frac{1}{\left|\mathrm{Stab}(1)\right|}\left|P_{\ket{\psi}}\left(0\right)-P_{\ket{\psi}}\left(1\right)\right| \\
    &=\frac{2}{3}\times0+\frac{1}{3}\times1
    \\&=\frac{1}{3}.
\end{align}
\end{proof}
Using this observation, we can then adapt techniques from Ref. \cite{aaronson2016complexitytheoretic} to prove the following Lemma:
\begin{lemma}\label{l:expected_TV_distance}
Let $\mu=\mathcal{U}_{\mathrm{Cl}(2^n)}$  or $\mu=\mu(n,d,\mathrm{Cl}(4))$. Then,
\begin{align}
  \underset{U\sim \mu}{\mathbb{E}}\left[ \mathrm{d}_{\mathrm{TV}}(P_U,\mathcal{U})\right] 
  &\geq \frac{1}{6}. \label{e:tv_expectation_cliff}
\end{align}
\end{lemma}
\begin{proof}
In Section 3.5 of Ref.~\cite{aaronson2016complexitytheoretic} it is proven that for Haar random nearest-neighbor circuits, the expected TV distance between the Born distribution $P_U$ and the uniform distribution $\mathcal{U}$ is lower bounded by $1/4$. In our notation, this can be written as
\begin{equation}\label{e:tv_expectation_haar}
    \underset{U\sim\mu(n,d,\mathrm{U}(4))}{\mathbb{E}}\left[ \mathrm{d}_{\mathrm{TV}}(P_U,\mathcal{U})\right] \geq \frac{1}{4},
\end{equation}
In their proof, the authors of Ref. \cite{aaronson2016complexitytheoretic} demonstrate that such a bound on the expected TV distance can be obtained using only one ingredient: the expected difference between the probability of measuring 0 and the probability of measuring 1 on a single-qubit Haar random state:
\begin{equation}
   \underset{U\sim\mathrm{U}(2)}{\mathbb{E}}[|P_U(0)-P_U(1)|] =  \underset{U\sim\mathrm{U}(2)}{\mathbb{E}}[\left| |\bra{0}U\ket{0}|^2-\bra{1}U\ket{0}|^2 \right|] =\frac{1}{2}. \label{e:expected_difference} 
\end{equation}
To leverage this expected difference between $P_U(0)$ and $P_U(1)$ from the single qubit case to a distribution over $n$ qubits the authors make use of the fact that a final random single-qubit gate can always be absorbed into a random circuit.

The proof of Lemma \ref{l:expected_TV_distance} - for both measures - proceeds completely analogously to the proof of Eq.~\eqref{e:tv_expectation_haar} given in Section 3.5 of Ref.~\cite{aaronson2016complexitytheoretic}, by exchanging Haar random gates with uniformly random Clifford gates, and using Lemma \ref{l:clifford_single_qubit} in place of Eq.~\eqref{e:expected_difference}. 
\end{proof}

Finally, using these ingredients, we can prove Lemma \ref{l:joint_prob_outside_ball}.

\begin{proof}(Lemma \ref{l:joint_prob_outside_ball})
For convenience, let us define $X:=\mathrm{d}_{\mathrm{TV}}(P_U,\mathcal{U}))$. By Markov's inequality, we have for any $\gamma >0$ that
\begin{align}
    \frac{1-\mathbb{E}(X)}{\gamma} &= \frac{\mathbb{E}[1-X]}{\gamma}\\
    &\geq \mathrm{Pr}\left[1-X \geq \gamma\right] \\
    &= \mathrm{Pr}\left[X\leq 1-\gamma \right] \\
    &= 1-\mathrm{Pr}\left[X\geq 1-\gamma \right],
\end{align}
from which it follows that
\begin{equation}
    \mathrm{Pr}\left[X\geq 1-\gamma \right] \geq 1 + \frac{\mathbb{E}(X)}{\gamma} -\frac{1}{\gamma}.
\end{equation}
Therefore, using the definition of $X$, along with the bound from Lemma \ref{l:expected_TV_distance}, and setting $\epsilon:=1-\gamma$, we obtain
\begin{equation}
    \underset{U\sim\mu}{\mathrm{Pr}} \left[ \mathrm{d}_{\mathrm{TV}}(P_U,\mathcal{U})  \geq \epsilon \right] \geq \frac{1/6-\epsilon}{1-\epsilon}.
\end{equation}
\end{proof}

\section{Proof of Lemma \ref{l:affine_sub_recovery}}\label{app:affine_recovery}

We provide in this section a proof for Lemma \ref{l:affine_sub_recovery}:

\affinesubrecoverylemma*

\noindent In order to prove this lemma we require some preliminary results and observations.

\begin{lemma}\label{l:subspace_trans} Given a vector subspace $U\subseteq\mathbb{F}^n_2$, and some $t\in\mathbb{F}^n_2$, let $A = t \oplus U$ be an affine subspace of $\mathbb{F}^n_2$. Then, for any $a \in A$, it is the case that $A\oplus a = U$.
\end{lemma}

\begin{proof}
Since $a\in A$, there exists some $u_1\in U$ such that $a=t\oplus u_1$. Using this one can see that
\begin{align}
    A\oplus a &= \{t\oplus u \oplus a\,|\, u \in U\} \\
    &= \{t\oplus u \oplus t\oplus u_1\,|\, u \in U\} \\
    & = \{u \oplus u_1\,|\, u \in U\}\\
    &= U,
\end{align}
where we have used that $t \oplus t = \mathbf{0}$ for all $t\in\mathbb{F}^n_2$, and that $U$ is closed under addition.
\end{proof}

\begin{lemma}\label{l:span_prob} Let $x_1,\ldots,x_k$ be $k$ vectors sampled uniformly at random from $\{0,1\}^n$. Let $P(k,n)$ denote the probability that the span of $x_1,\ldots,x_k$ is $\mathbb{F}^n_2$, i.e. 
\begin{equation}
    P(k,n) = \mathrm{Pr}[\mathrm{span}\{x_1,\ldots x_k\} = \mathbb{F}^n_2].
\end{equation}
Then, $P(k,n) = 0$ for all $k<n$, and for $k\geq n$ 
\begin{equation}
    P(k,n) = \prod_{j=0}^{n-1}\left(1-2^{j-k}\right).
\end{equation}
\end{lemma}
\begin{proof}
As $\mathbb{F}^n_2$ is $n$-dimensional the fact that $P(k,n)=0$ for all $k<n$ is immediate. For $k\geq n$, let us consider the $k\times n$ binary matrix with $x_1,\ldots,x_k$ as rows. $P(k,n)$ is then the probability that this matrix has row-rank $n$, which is equal to the probability that this matrix has column-rank $n$. As such, we proceed by calculating the probability that a uniformly drawn $k\times n$ binary matrix has full column-rank. To do this we start by noting that, for $k\geq n$, the number of full column-rank binary $k\times n$ matrices is given by
\begin{equation}
    F(k,n) = \prod_{j = 0}^{n-1}\left(2^k - 2^j\right).
\end{equation}
To see this, note that there are $2^{k}-1$ choices to pick a linearly independent vector for the first column (any vector except the all zero vector). Having fixed the first column, there are then $2^k - 2^1$ choices for a second column which is linearly independent from the first. In general, if the first $j$ columns are linearly independent, there are $2^k-2^j$ choices for the $j+1$'th column.

To complete the proof, we note that the total possible number of $k\times n$ binary matrices is $2^{kn}$, and therefore the probability of drawing a full-column rank (and therefore row-rank $n$) $k\times n$ binary matrix is 
\begin{equation}
    P(k,n) = \frac{F(k,n)}{2^{kn}} = \prod_{j = 0}^{n-1}\left(1-2^{j-k}\right).
\end{equation}
\end{proof}

\begin{corollary}\label{c:prob_rec}
Let $U$ be an $m$-dimensional subspace of $\mathbb{F}^n_2$, and let $x_1,\ldots,x_k$ be $k\geq m$ vectors sampled uniformly at random from $U$. Then
\begin{equation}
    \mathrm{Pr}[\mathrm{span}\{x_1,\ldots,x_k\} = U] \geq 1-2^{m-k}.
\end{equation}
\end{corollary}
\begin{proof}
Recall that any such subspace $U$ is isomorphic to $\mathbb{F}^m_2$, and therefore
\begin{equation}
    \mathrm{Pr}[\mathrm{span}\{x_1,\ldots,x_k\} = U] = P(k,m).
\end{equation}
Using Lemma \ref{l:span_prob} we then have that
\begin{align}
    P(k,m) &= \prod_{j = 0}^{m-1}\left(1-2^{m-k}\right) \\
    &\geq 1 - 2^{-k}\sum_{j = 0}^{m-1}2^j \\
    &= 1 - 2^{m-k} - 2^{-k} \hspace{5em} \left(\text{via} \sum_{j = 0}^{m-1}2^j = 2^n -1\right) \\
    &\geq 1 - 2^{m-k}.
\end{align}
\end{proof}

\noindent Using these ingredients we can now provide a proof for Lemma \ref{l:affine_sub_recovery}, via Algorithm \ref{alg:aff_rec}.

\begin{algorithm}[H]
  \caption{Affine subspace recovery algorithm
    \label{alg:aff_rec}}
  \begin{algorithmic}[1]
  	\Statex Given $\delta\in(0,1)$ and access to $\sample(U_A)$ for some affine subspace $A\subseteq{\mathbb{F}^n_2}$,
    \Statex
    \State Set $k:=n + \lceil\log(1/\delta)\rceil$.
    \State Query $k$ many times and obtain $\{x_1,\ldots,x_k\}$.
    \State Transform the samples $x_1,\ldots,x_k$ to $\tilde{x}_1,\ldots,\tilde{x}_k$ via $\tilde{x}_i = x_i\oplus x_1$. \label{al:transform}
    \State Use Gaussian elimination to determine from $\tilde{x}_1,\ldots,\tilde{x}_k$ a maximal linearly independent subset of vectors $V :=\{\tilde{x}_{i_1},\ldots,\tilde{x}_{i_m}\}$.\label{al:lin_find}
    \State Form the full rank $n\times m$ matrix $\mathbf{R}$ by placing vectors from $V$ as columns.
    \State Output $(\mathbf{R},x_1)$.
  \end{algorithmic}
\end{algorithm}

\begin{proof}(Lemma \ref{l:affine_sub_recovery}) We claim that the the affine subspace recovery algorithm described as Algorithm \ref{alg:aff_rec} satisfies all required properties. To see this, we start by assuming that the affine subspace $A$ is described by some tuple $(\mathbf{R}',t')$, and we denote by $U :=\{\mathbf{R}'b\,|\,b\in\mathbb{F}^n_2\}$ the subspace defined by $\mathbf{R}'$, such that $A = t\oplus U$.  We then note that, as a consequence of Lemma \ref{l:subspace_trans}, Step~\ref{al:transform} of Algorithm \ref{alg:aff_rec} transforms the $k$ samples from $U_A$ into $k$ samples $\tilde{x_1},\ldots,\tilde{x}_k$ drawn uniformly from the vector subspace $U$. From Corollary \ref{c:prob_rec} we then have that
\begin{align}
    \mathrm{Pr}[\mathrm{span}\{\tilde{x_1},\ldots,\tilde{x}_k\}=U] &\geq 1 - 2^{m-k} \\
    &\geq 1 - 2^{n-k} \hspace{10em} (\text{via } n\geq m)\\
    &\geq 1 - 2^{n - (n+\log(1/\delta))} \hspace{6em} (\text{via } k:=n + \lceil\log(1/\delta)\rceil) \\
    &=1 - \delta.
\end{align}
Hence, with probability at least $1-\delta$, the columns of $\mathbf{R}$ provide a basis for $U$. Let's assume this is the case, and consider the affine subspace $B:=x_1\oplus U$ - i.e. the affine subspace described by the tuple $(\mathbf{R},x_1)$. Noting that $x_1 = t \oplus v$ for some $v\in U$ (as a consequence of the fact that $x_1\in A=t\oplus U$) we have that
\begin{align}
    B &= \{x_1 \oplus u \,|\, u\in U\}\\
    &=\{t\oplus v\oplus u\,|\, u\in U\}\\
    &=\{t\oplus u'\,|\, u'\in U\}\\
    &= A,
\end{align}
where the third line follows from the fact that $U$ is closed under addition. As a result, we see that indeed, with probability at least $1-\delta$, the tuple $(\mathbf{R},t)$ describes the affine subspace $A$. The claimed efficiency of Algorithm \ref{alg:aff_rec} follows immediately from the efficiency of Gaussian elimination, and the fact that only $k:=n+\lceil\log(1/\delta)\rceil$ queries are required.
\end{proof}

\bibliographystyle{alpha}
\bibliography{literature.bib}

\end{document}